\documentclass[pra,showpacs,nofootinbib]{revtex4-1}
\usepackage[latin1]{inputenc}
\usepackage{amsmath,amssymb,amscd,bbm,epsfig,bm}
\usepackage{graphicx}
\usepackage{color}
\newcommand{\assign}{:=}
\newcommand{\bigintlim}{\int}
\newcommand{\bignone}{\,}

\newcommand{\mathd}{\mathrm{d}}
\newcommand{\mathe}{\mathrm{e}}

\newcommand{\tmem}[1]{{\em #1\/}}
\newcommand{\tmmathbf}[1]{\ensuremath{\boldsymbol{#1}}}
\newcommand{\tmop}[1]{\ensuremath{\operatorname{#1}}}
\newcommand{\tmtextbf}[1]{{\bfseries{#1}}}
\newcommand{\tmtextit}[1]{{\itshape{#1}}}

\newcommand{\tmscript}[1]{\scriptstyle{#1}}

\begin{document}

\title{Quantum master equation for collisional dynamics of massive particles with internal degrees of
freedom}
\author{Andrea \surname{Smirne}${}^{a,b}$}
\email{andrea.smirne@unimi.it}
\author{Bassano \surname{Vacchini}${}^{a,b}$}
\email{bassano.vacchini@mi.infn.it}
\affiliation{\mbox {${}^a$Universit{\`a} degli Studi di Milano, Dipartimento di
Fisica, Via Celoria 16, I-20133 Milano, Italy}\\
\mbox{${}^b$INFN, Sezione di Milano, Via Celoria 16, I-20133 Milano, Italy}
}
\date{\today}

\begin{abstract}
We address the microscopic derivation of a quantum
master equation in Lindblad form for the dynamics of a massive test
particle with internal degrees of freedom, interacting through
collisions with a background ideal gas. When either internal or centre
of mass degrees of freedom can be treated classically, previously
established equations are obtained as special cases. If in an
interferometric setup the internal degrees of freedom are not detected
at the output, the equation can be recast in the form of a generalized
Lindblad structure, which describes non-Markovian effects. The effect
of internal degrees of freedom on centre of mass decoherence is
considered in this framework.
\end{abstract}
\maketitle

\section{Introduction\label{sec:intro}}

In recent times major advances in the experimental techniques have led to the
realization of experiments in which quantum systems in a single particle
regime are studied under their controlled interaction with some environment. A
paradigmatic example in this context is given by the motion of a massive test
particle in an interferometric setup, which gives rise to interference fringes
as typical quantum signatures. When the coupling with the environment becomes
of relevance, such interference fringes are gradually washed out, and a
classical dynamics is eventually recovered. This phenomenon goes under the
name of decoherence {\cite{Joos2003,Schlosshauer2007,Breuer2007}}. Its
understanding and theoretical description require on the one hand a control
over the environment, on the other hand a microscopic model for the
interaction and the ensuing dynamics.

For the case of a tracer particle immersed in a dilute gas such a microscopic
description has been obtained considering the centre of mass degrees of
freedom only. The reduced dynamics is given by a master equation in Lindblad
form which has been called quantum linear Boltzmann equation, since it
provides the natural quantum counterpart of the classical linear Boltzmann
equation (see {\cite{Vacchini2009a}} for a recent review and references
therein). The microscopic input is given by the complex scattering amplitudes
describing the collisions between gas and test particle, while the gas is
characterized by its density and momentum distribution. In this paper we
consider an extension of this result, which includes internal degrees of
freedom of the tracer particle. The microscopic derivation is performed along
the lines of a general strategy for the derivation of Markovian master
equations, which relies on a scattering description of the interaction events
{\cite{Hornberger2007b}}. Besides the gas properties, this approach takes as
basic input the multichannel complex scattering amplitudes, which describe the
influence of the internal states on the scattering events. Indeed, when the
scattering cross section does not only depend on the relative motional state
between tracer and gas particle, such an extension becomes mandatory in order
to correctly describe the dynamics. According to the Markovian approximation,
the obtained master equation is in Lindblad form. This derivation confirms the
structure of the dissipative term, which has been heuristically obtained in
{\cite{Vacchini2008a}}, further determining the coherent contribution to the
dynamics due to forward scattering. The latter becomes relevant in the
determination of the index of refraction for matter waves. When either type of
degrees of freedom can be described in classical terms, a Markovian quantum
classical master equation is obtained. Such a result, corresponding to a
classical treatment of the motional degrees of freedom, has been considered in
{\cite{Kryszewski2006a}}. In that context the name Bloch-Boltzmann equation
was proposed for the equation, since for a two-level system an extension of
the optical Bloch equations to include a Boltzmann-like collision term is
obtained. In the same spirit, the name quantum Bloch-Boltzmann equation can be
used to indicate a master equation, which gives a quantum description of both
internal and centre of mass state.

An interesting situation appears when in the final detection the internal
state of the test particle is not resolved at the output of the
interferometer. In this case the internal degrees of freedom become part of
the environment. Then a non-Markovian dynamics for the motional state appears,
which can be described in terms of a coupled set of Lindblad equations for the
unnormalized statistical operators corresponding to specific internal
channels. This type of non-Markovian dynamics can be considered as a
generalized non-Markovian Lindblad structure. It arises as a mean over a
classical index, which can take place e.g. as a consequence of the interaction
with a structured reservoir {\cite{Budini2006a,Breuer2007a,Vacchini2008a}}.
This situation is here considered in the study of the loss of visibility of
the interference fringes in an interferometric setup. The ensuing decoherence
effect is generally not described as an exponential loss of visibility
depending on the strength of the interaction, as in the usual Markovian case.

The paper is organized as follows. In Sect.~\ref{sec:ms} we consider
the expression of the master equation, pointing to the main steps
necessary for its derivation and putting into evidence the microscopic
quantities determining its explicit form. A detailed microscopic
derivation of the master equation is performed in
Appendix~\ref{sec:micr-deriv-mast}.  The master equation is given both
in terms of matrix elements of the statistical operator in the
momentum and internal energy eigenstates basis, as well as an explicit
operator expression, which makes its Lindblad structure manifest. This
also allows to easily recover under suitable limits previously
considered master equations, which describe either only one of the two
kind of degrees of freedom or a hybrid quantum classical description
of both. In Sect.~\ref{sec:decoh} we show how the interplay between
internal and motional states can influence the visibility in an
interferometric setup for the study of decoherence, leading to a
non-Markovian behaviour in the reduction of the visibility of the
interference fringes.

\section{The master equation for a test particle with internal degrees of
freedom\label{sec:ms}}

We first consider the key ingredients and steps which lead to obtain
the master equation describing the collisional dynamics of a test
particle immersed in a structureless background gas, keeping the
internal degrees of freedom of the particle into account.  The task of
a full microscopic derivation will be accomplished in
Appendix~\ref{sec:micr-deriv-mast}, relying on a method recently
introduced for the derivation of Markovian master equations, which has
been called monitoring approach
{\cite{Hornberger2006b,Hornberger2007b,Hornberger2007a,Hornberger2008a}}.
In the monitoring approach the reduced dynamics of a system in contact with
some environment is obtained describing their interaction by means of
scattering theory. The building blocks in such a formulation of the open
system dynamics are therefore the S-matrix characterizing the single
interaction events and the rate of collisions. Both quantities are given by
operators on the tensor product Hilbert space of system and environment, which
we shall denote by $\mathsf{\mathsf{S}} = \mathsf{I} + i \mathsf{T}$ and
$\mathsf{\Gamma}$ respectively. The operator nature of these quantities is
crucial in order to keep the gas and test particle state into account in the
dynamic description of the collisional interaction. The Markovian master
equation for the reduced dynamics is obtained by assuming the various
collisions as independent, so that their effect cumulates according to the
state dependent scattering rate, and taking the trace over the environmental
degrees of freedom.

\subsection{Expression of the master equation}

The formal expression of the master equation reads {\cite{Hornberger2007b}}
\begin{eqnarray}
  \frac{\mathd}{\tmop{dt}} \rho & = & \frac{1}{i \hbar}  \left[ \mathsf{H},
  \rho \right] +\mathcal{L} \rho +\mathcal{R} \rho  \label{eq:qlbefinal}
\end{eqnarray}
where $\mathsf{H}$ is the free Hamiltonian and $\rho$ is the statistical
operator of the system. For the case at hand the free
Hamiltonian of the system is given by
\begin{eqnarray}
  \mathsf{H} & = & \frac{\mathsf{P}^2}{2 M} \otimes \sum_i \hbar \omega_i |i
  \rangle \langle i|,  \label{eq:ho}
\end{eqnarray}
where $\mathsf{P}$ is the momentum operator of the test particle, $M$ its
mass and $\left\{ |i \rangle \right\}_{i = 1, \ldots, n}$ the basis of energy
eigenstates in $\mathbbm{C}^n$. The superoperators appearing at r.h.s. of
Eq.~(\ref{eq:qlbefinal}) are defined according to
\begin{eqnarray}
  \mathcal{L} \rho & = & \text{Tr}_{\tmop{gas}} \left( \mathsf{T}
  \mathsf{\Gamma}^{1 / 2} \left[ \rho \otimes \rho_{\text{gas}} \right]
  \mathsf{\Gamma}^{1 / 2}  \mathsf{T}^{\dag} \right) \nonumber\\
  &  & - \frac{1}{2} \text{Tr}_{\tmop{gas}} \left( \mathsf{\Gamma}^{1 / 2}
  \mathsf{T}^{\dag} \mathsf{T} \mathsf{\Gamma}^{1 / 2} \left[ \rho \otimes
  \rho_{\text{gas}} \right] \right) \nonumber\\
  &  & - \frac{1}{2} \text{Tr}_{\tmop{gas}} \left( \left[ \rho \otimes
  \rho_{\text{gas}} \right] \mathsf{\Gamma}^{1 / 2} \mathsf{T}^{\dag}
  \mathsf{T} \mathsf{\Gamma}^{1 / 2} \right)  \label{eq:ms}
\end{eqnarray}
and
\begin{eqnarray}
  \mathcal{R} \rho & = & i \text{Tr}_{\tmop{gas}} 
\left( 
\left[
\mathsf{\Gamma}^{1 /
  2} \tmop{Re} \left( \mathsf{T} \right) \mathsf{\Gamma}^{1 / 2},  \rho
  \otimes \rho_{\text{gas}} 
\right]
\right)  \label{eq:r}
\end{eqnarray}
respectively, where $\rho_{\text{gas}}$ is the single particle
statistical operator describing the gas environment.  Note that the
operators $\mathcal{L}$ and $\mathcal{R}$ arise by acting with an
operator in Lindblad form on a state of system plus gas in factorized
form, further taking the partial trace with respect to the gas. While
this operation is formally legitimate, and guarantees preservation of
trace and Hermiticity of the statistical operator describing the test
particle, it is generally not true that the resulting dynamics for the
reduced system only is given by a master equation in Lindblad form,
thus granting complete positivity and describing a well-defined
Markovian dynamics. Indeed this step involves further approximations,
which depend in a crucial way on details of system and interaction. It
is well known that by taking the partial trace with respect to the
unitary evolution of the overall system one can obtain a Markovian
dynamics only if further hypotheses hold. This remains true for the
case at hand, despite the fact that important approximations have
already been introduced in replacing the Hamiltonian dynamics for
system plus gas with a Lindblad operator only specified by
$\mathsf{T}$ and $\mathsf{\Gamma}$. The actual proof that a Markovian
dynamics applies to the situation of interest and the specific
expression of the superoperators appearing in Eq.~(\ref{eq:qlbefinal})
is obtained through the microscopic calculations performed in
Appendix~\ref{sec:micr-deriv-mast}.

Relying on the results of Appendix~\ref{incoh} we write
the following expression for the contributions in Eq.~(\ref{eq:ms})  in the momentum and channel
basis $\{|\tmmathbf{P}, i \rangle \}$
\begin{eqnarray}
\nonumber  \langle \tmmathbf{P}, i|\mathcal{L} \rho |\tmmathbf{P}', k \rangle & = &
  \sum_{j l} \int d\tmmathbf{Q}\left[ \langle \tmmathbf{P}-\tmmathbf{Q}, j|
  \rho |\tmmathbf{P}' -\tmmathbf{Q}, l \rangle M^{j l}_{i k} \left(
  \tmmathbf{P}, \tmmathbf{P}' ; \tmmathbf{Q} \right) \right. \nonumber\\
  & & - \frac{1}{2} \langle \tmmathbf{P}, j| \rho |\tmmathbf{P}', k \rangle
  M^{j i}_{l l} (\tmmathbf{P}+\tmmathbf{Q}, \tmmathbf{P}+\tmmathbf{Q};
  \tmmathbf{Q}) \nonumber\\
&  &- \left.  \frac{1}{2} \langle \tmmathbf{P}, i| \rho |\tmmathbf{P}', l
  \rangle M^{k l}_{j j} (\tmmathbf{P}' +\tmmathbf{Q}, \tmmathbf{P}'
  +\tmmathbf{Q}; \tmmathbf{Q}) \right] \label{eq:rif2}
\end{eqnarray}
where the complex rate functions $M^{j l}_{i k} \left( \tmmathbf{P},
   \tmmathbf{P}' ; \tmmathbf{Q} \right)$
are given by
\begin{eqnarray}
  M^{j l}_{i k} \left( \tmmathbf{P}, \tmmathbf{P}' ; \tmmathbf{Q} \right) & =
  & \chi^{jl}_{ik} \int_{\tmmathbf{Q}_{\perp}} d\tmmathbf{p}\,\,L_{i j} \left(
  \tmmathbf{p}, \tmmathbf{P}-\tmmathbf{Q}; \tmmathbf{Q} \right) L_{k l}^{\ast}
  \left( \tmmathbf{p}, \tmmathbf{P}' -\tmmathbf{Q}; \tmmathbf{Q} \right),
  \label{eq:fact}
\end{eqnarray}
with the $\chi^{jl}_{ik}$ a notational shorthand to indicate that the
contribution is different from zero only for $\mathcal{E}_{i j}
=\mathcal{E}_{k l}$, where $\mathcal{E}_{k j} = E_k - E_j $ denotes the
difference in energy between internal states, while the
$\tmmathbf{p}$-integration is restricted to the plane
$\tmmathbf{Q}_{\perp} = \{ \tmmathbf{p} \in \mathbbm{R}^3 :
\tmmathbf{p} \cdot \tmmathbf{Q}= 0 \}$. The functions $L_{i j} \left(
   \tmmathbf{p}, \tmmathbf{P}; \tmmathbf{Q} \right)$ are defined
according to
\begin{eqnarray}
  L_{i j} \left( \tmmathbf{p}, \tmmathbf{P}; \tmmathbf{Q} \right) & = &
  \sqrt{\frac{n_{\tmop{gas}} m}{m^2_{\ast} Q}} \sqrt{\mu \left(
  \tmmathbf{p}_{\perp} + \frac{m}{M} \tmmathbf{P}_{\parallel} + \left( 1 +
  \frac{m}{M} \right) \frac{\tmmathbf{Q}}{2} + \frac{\mathcal{E}_{i j}}{Q^2 /
  m} \tmmathbf{Q} \right)}\nonumber\\
  & &\times  f_{i j} \left( \text{rel} \left( \tmmathbf{p}_{\perp},
  \tmmathbf{P}_{\perp} \right) - \frac{\tmmathbf{Q}}{2} + \frac{\mathcal{E}_{i
  j}}{Q^2 / m_{\ast}} \tmmathbf{Q}, \text{rel} \left( \tmmathbf{p}_{\perp},
  \tmmathbf{P}_{\perp} \right) + \frac{\tmmathbf{Q}}{2} + \frac{\mathcal{E}_{i
  j}}{Q^2 / m_{\ast}} \tmmathbf{Q} \right), \nonumber \\ \label{eq:L}
\end{eqnarray}
where $\mu (\tmmathbf{p})$ denotes the stationary gas distribution,
$n_{\tmop{gas}}$ is the density of the gas, $m$ the mass of the gas
particles, $m_{\ast} = m M / (M + m)$ the reduced mass, and $f_{k j}
(\tmmathbf{p}_f, \tmmathbf{p}_i)$ denote the multichannel complex
scattering amplitudes, which depend on the microscopic interaction
potential and describe scattering from an initial momentum
$\tmmathbf{p}_i$ and internal state $j$ to a final state with momentum
$\tmmathbf{p}_f$ and internal state $k$.  Moreover, $
\tmmathbf{P}_{\perp}$ and $ \tmmathbf{P}_{\parallel}$ indicate,
respectively, the perpendicular and the parallel component of the
momentum $ \tmmathbf{P}$ with respect to the vector $ \tmmathbf{Q}$;
while $\tmop{rel} (\tmmathbf{p}, \tmmathbf{P}) \equiv \left( m_{\ast}
   / m \right) \tmmathbf{p}- \left( m_{\ast} / M \right) \tmmathbf{P}$
is the the relative momentum between the gas particle momentum
$\tmmathbf{p}$ and the test particle momentum $\tmmathbf{P}$.

Exploiting these results we can easily write the master equation
Eq.~(\ref{eq:qlbefinal}) directly in operator form. In fact using
the functions $L_{i j} \left( \tmmathbf{p}, \tmmathbf{P}; \tmmathbf{Q}
\right)$ let us introduce the following family of jump operators
\begin{eqnarray}
  \mathsf{L}_{\tmmathbf{Q}, \tmmathbf{p}, \mathcal{E}} & = &
  \mathe^{i\tmmathbf{Q} \cdot \mathsf{X} / \hbar}
  \sum_{\underset{\mathcal{E}_{ij} = \mathcal{E}}{ij}} L_{i j} \left(
  \tmmathbf{p}, \mathsf{P} ; \tmmathbf{Q} \right) \otimes \mathsf{E}_{ij},
  \label{eq:lindbladop}
\end{eqnarray}
where $\mathsf{X}$ and $\mathsf{P}$ are position and momentum operators of the
test particle, while the operators $\mathsf{E}_{i j} = |i \rangle \langle j|$
act on the internal degrees of freedom only, since $|i \rangle$ denotes the
energy eigenstate with eigenvalue $\hbar \omega_i$, and the exponential factor
describes momentum exchanges according to $\exp \left( i \mathsf{X} \cdot
\tmmathbf{Q}/ \hbar \left) |\tmmathbf{P} \rangle = |\tmmathbf{P}+\tmmathbf{Q}
\rangle \right. \right.$. Note that the functions $L_{i j} \left(
\tmmathbf{p}, \tmmathbf{P}; \tmmathbf{Q} \right)$ essentially depend on the
scattering amplitudes and the momentum distribution of the gas, thus keeping
into account all the details of the collisional interaction. These expressions
appear operator-valued in the master equation, being evaluated for
$\tmmathbf{P} \rightarrow \mathsf{P}$, so as to take into account the actual
momentum of the colliding test particle.
The incoherent contribution $ \mathcal{L}$ in Eq.~(\ref{eq:qlbefinal})
finally reads
\begin{eqnarray}
  \mathcal{L} \rho & = & \sum_{\mathcal{E}} \int d\tmmathbf{Q}
  \int_{\tmmathbf{Q}_{\perp}} d\tmmathbf{p} \left( \mathsf{L}_{\tmmathbf{Q},
  \tmmathbf{p}, \mathcal{E}} \rho \mathsf{L}^{\dag}_{\tmmathbf{Q},
  \tmmathbf{p}, \mathcal{E}} - \frac{1}{2} \left\{
  \mathsf{L}^{\dag}_{\tmmathbf{Q}, \tmmathbf{p}, \mathcal{E}}
  \mathsf{L}_{\tmmathbf{Q}, \tmmathbf{p}, \mathcal{E}}, \rho \right\}
\right).
\label{eq:diag}
\end{eqnarray}
The superoperator $\mathcal{R}$ of Eq.~(\ref{eq:r}) on its turn
according to Appendix~\ref{coh} amounts
to the commutator with an effective Hamiltonian given by
\begin{eqnarray}
  \mathsf{H}_n & = &  \text{$- 2 \pi \hbar^2 \frac{n_{\tmop{gas}}}{m_{\ast}}
  \sum_{\underset{\mathcal{E}_{ij} = 0}{ij}} \bignone \int d\tmmathbf{p}_0 \mu
  (\tmmathbf{p}_0)$}\text{Re}  \left[ f_{i j} \left( \text{rel} \left(
  \tmmathbf{p}_0, \mathsf{P} \right), \text{rel} \left( \tmmathbf{p}_0,
  \mathsf{P}  \right)  \right)  \right] \otimes \mathsf{E}_{ij}.
  \label{eq:hn}
\end{eqnarray}
Using the alternative expression given by Eq.~(\ref{eq:L2}) to define $L_{i j} \left(
\tmmathbf{p}, \tmmathbf{P}; \tmmathbf{Q} \right)$, it is immediately seen that
the incoherent term of this master equation confirms the result heuristically
obtained in {\cite{Vacchini2008a}}. In the latter reference this equation has
also been termed quantum Bloch-Boltzmann equation in that it provides a
quantum description of both motional and internal degrees of freedom, thus
extending the result of {\cite{Alicki2003a}}, where the centre of mass degrees
of freedom were treated classically and the name Bloch-Boltzmann equation was
used. These names should not confuse the reader. Indeed, only for the case of
an atom in a two-level approximation undergoing a collisional dynamics this
equation refers to an extension of the optical Bloch equations with a
Boltzmann collision term.

\subsection{Limiting forms}

As a compatibility check of the master equation derived in
Appendix~\ref{sec:micr-deriv-mast}, and in order to make contact with previous
work, we will now show how in suitable limits it recovers already
known equations. Since the equation describes the quantum dynamics of
a test particle with both internal and translational degrees of
freedom immersed in a dilute gas, natural limiting situations appear
considering a structureless test particle or an immobile system. These
situations correspond to the quantum linear Boltzmann equation
{\cite{Hornberger2008a}}, and to the master equation for an immobile
system interacting through collisions with a background gas
{\cite{Dumcke1985a,Hornberger2007b}}. Another natural limit consists
in a hybrid quantum classical description, in which either internal or
centre of mass degrees of freedom are treated classically. The master
equation corresponding to this last case has already been considered
in {\cite{Alicki2003a}}. A classical treatment of both kinds of
degrees of freedom leads to the master equation for a classical Markov
process, with a probability density depending on both a discrete and a
continuous index.

\paragraph{Quantum linear Boltzmann equation \label{boltzmann}}

If the internal degrees of freedom can be disregarded the sum in
Eq.~(\ref{eq:diag}) has a single non vanishing contribution, so that instead
of the multichannel scattering amplitudes $f_{i j} (\tmmathbf{p}_f,
\tmmathbf{p}_i)$ there is a single amplitude which can be indicated as $f
(\tmmathbf{p}_f, \tmmathbf{p}_i)$. The incoherent term in the master equation
reduces to
\begin{eqnarray}
  \mathcal{L} \rho & = & \int d\tmmathbf{Q} \int_{\tmmathbf{Q}_{\perp}}
  d\tmmathbf{p} \left( \mathsf{L}_{\tmmathbf{Q}, \tmmathbf{p}} \rho
  \mathsf{L}^{\dag}_{\tmmathbf{Q}, \tmmathbf{p}} - \frac{1}{2} \left\{
  \mathsf{L}^{\dag}_{\tmmathbf{Q}, \tmmathbf{p}}  \mathsf{L}_{\tmmathbf{Q},
  \tmmathbf{p}}, \rho \right\} \right)  \label{eq:qlbe}
\end{eqnarray}
with
\begin{eqnarray*}
  \mathsf{L}_{\tmmathbf{Q}, \tmmathbf{p}} & = & \sqrt{\frac{n_{\tmop{gas}}
  m}{m^2_{\ast} Q}} \mathe^{i\tmmathbf{Q} \cdot \mathsf{X} / \hbar} \sqrt{\mu
  \left( \tmmathbf{p}_{\perp} + \frac{m}{M} \mathsf{P}_{\parallel} + \left( 1
  + \frac{m}{M} \right) \frac{\tmmathbf{Q}}{2} \right)}\nonumber\\
  & &\times f \left( \text{rel}
  \left( \tmmathbf{p}_{\perp}, \mathsf{P}_{\perp} \right) -
  \frac{\tmmathbf{Q}}{2}, \text{rel} \left( \tmmathbf{p}_{\perp},
  \mathsf{P}_{\perp} \right) + \frac{\tmmathbf{Q}}{2} \right),
\end{eqnarray*}
while the Hamiltonian term reads $\mathsf{H}_0 + \mathsf{H}_n$, where
$\mathsf{H}_0 = \mathsf{P}^2 / 2 M$ and
\begin{eqnarray*}
  \mathsf{H}_n & = & - \frac{2 \pi \hbar^2 n_{\tmop{gas}}}{m_{\ast}}  \int
  d\tmmathbf{p}_0 \mu (\tmmathbf{p}_0) \text{Re}  \left[ f \left( \text{rel}
  \left( \tmmathbf{p}_0, \mathsf{P} \right), \text{rel} \left( \tmmathbf{p}_0,
  \mathsf{P} \right)  \right)  \right]
\end{eqnarray*}
takes into account the energy shift due to forward scattering. This result
complies with the quantum linear Boltzmann equation obtained in
{\cite{Hornberger2008a}}, whose properties have been discussed in detail in
{\cite{Vacchini2009a}}.

\paragraph{Immobile tracer particle \label{bloch}}

We now consider the opposite situation, corresponding to an infinitely massive
test particle, so that the dynamics of the translational degrees of freedom
can be neglected. To consider this limit it is convenient to come back to the
expression Eq.~(\ref{eq:rif2}) of the quantum master equation in terms
of the complex rate functions $M^{j l}_{i
k} \left( \tmmathbf{P}, \tmmathbf{P}' ; \tmmathbf{Q} \right)$, which
in the limit $M \rightarrow \infty$ when integrated over $\tmmathbf{Q}$ reduce to
\begin{eqnarray}
  M^{j l}_{i k} & = & \frac{n_{\tmop{gas}}}{m^2} \chi^{jl}_{ik}   \int
  d\tmmathbf{p} \int d\tmmathbf{p}_0 \delta \left( \frac{\tmmathbf{p}^2
  -\tmmathbf{p}^2_0}{2 m} + E_i - E_j \right) f_{i j} (\tmmathbf{p},
  \tmmathbf{p}_0) f^{\ast}_{k l} (\tmmathbf{p}, \tmmathbf{p}_0),
  \label{eq:Mimmombile}
\end{eqnarray}
where no dependence on the test particle's momentum is left. The matrix
elements of the incoherent part of the quantum Bloch-Boltzmann equation are
therefore now given by
\begin{eqnarray}
  \langle i|\mathcal{L} \rho | k \rangle & = & \sum_{j l} \left( \langle j|
  \rho |l \rangle M^{j l}_{i k} - \frac{1}{2} \langle j| \rho | k \rangle M_{l
  l}^{j i} - \frac{1}{2} \langle i| \rho |l \rangle M_{j j}^{k l}  \right),
  \label{eq:coherimmobile}
\end{eqnarray}
while the coherent part corresponds to a effective Hamiltonian 
whose matrix elements in the energy eigenbasis are given by $E^{ij}_n = - 2 \pi \hbar^2 ({n_{\tmop{gas}}}/{m_{\ast}})
  \chi^{jk}_{ik} \bignone \int d\tmmathbf{p}_0 \mu (\tmmathbf{p}_0) \text{Re}
  \left[ f_{i j} \left( \tmmathbf{p}_0, \tmmathbf{p}_0 \right) \right]$,
thus confirming the result obtained in {\cite{Hornberger2007b}} for
the case of a non degenerate Hamiltonian.

\paragraph{Quantum classical description}

The limiting expressions of the quantum Bloch-Boltzmann equation, obtained
when either the internal or the translational degrees of freedom can be
treated as a classical label, correspond to hybrid quantum classical
descriptions, which naturally arise when
decoherence affects on different time scales the two kind of degrees of
freedom.


When the centre of mass degrees of freedom can be treated classically it
is convenient to introduce the classical rates
\begin{eqnarray}
  M^{j l}_{i k} \left( \tmmathbf{P}+\tmmathbf{Q}; \tmmathbf{Q} \right) & : = &
  M^{j l}_{i k} \left( \tmmathbf{P}+\tmmathbf{Q}, \tmmathbf{P}+\tmmathbf{Q};
  \tmmathbf{Q} \right)  \label{eq:clarate}
\end{eqnarray}
with $M^{j l}_{i k} \left( \tmmathbf{P}, \tmmathbf{P}' ; \tmmathbf{Q} \right)$
as in Eq.~(\ref{eq:fact}), so that the semiclassical Bloch-Boltzmann equation reads
\begin{eqnarray}
  \frac{\mathd}{\tmop{dt}} \rho (\tmmathbf{P}) & = & \frac{1}{i \hbar} \left[
  \sum_i \hbar \omega_i |i \rangle \langle i| + \mathsf{H}_n (\tmmathbf{P}),
  \rho (\tmmathbf{P}) \right] + \sum_{\tmscript{\begin{array}{c}
    ijkl
  \end{array}}} \bigintlim \mathd \tmmathbf{Q} \left[ M^{jl}_{ik} \left(
  \tmmathbf{P}; \tmmathbf{Q} \right) \mathsf{E}_{ij} \rho
  (\tmmathbf{P}-\tmmathbf{Q}) \mathsf{E}^{^{\dag}}_{kl} \right. \nonumber\\
  &  & \left. - \frac{1}{2} M^{jl}_{ik} \left( \tmmathbf{P}+\tmmathbf{Q};
  \tmmathbf{Q} \right) \left\{ \mathsf{E}^{^{\dag}}_{kl} \mathsf{E}_{ij}, \rho
  (\tmmathbf{P}) \right\} \right],  \label{eq:ratebis}
\end{eqnarray}
where $\mathsf{H}_n (\tmmathbf{P})$ is obtained from Eq.~(\ref{eq:hn}) with the
replacement $\mathsf{P} \rightarrow \tmmathbf{P}$, and $\rho
(\tmmathbf{P})$ denotes a collection of trace class operators in
$\mathbbm{C}^n$ normalized according to $\bigintlim \mathd \tmmathbf{P} \tmop{Tr}_{\mathbbm{C}^n} \text{$\rho
  (\tmmathbf{P})$}=1$.


If the classical approximation applies for the internal degrees of
freedom the incoherent term of the master equation giving a quantum
description of the translational dynamics only takes the form
\begin{eqnarray}
  \frac{\mathd}{\tmop{dt}} \rho_i & = & \frac{1}{i \hbar} \left[
  \frac{\mathsf{P}^2}{2 M} + \mathsf{H}^i_n, \rho_i \right] + \sum_j
  \bigintlim \mathd \tmmathbf{Q} \int_{\tmmathbf{Q}^{\bot}} \bignone \mathd
  \tmmathbf{p} \left[ 
\mathe^{i\tmmathbf{Q} \cdot \mathsf{X} / \hbar} L_{i j} \left( \tmmathbf{p},
   \mathsf{P} ; \tmmathbf{Q} \right)
\rho_j 
L_{i j} \left( \tmmathbf{p},
   \mathsf{P} ; \tmmathbf{Q} \right)^{\dag} 
\mathe^{-i\tmmathbf{Q} \cdot \mathsf{X} / \hbar}
\right. \nonumber\\
  &  & \left. - \frac{1}{2} \left\{ 
L_{i j} \left( \tmmathbf{p},
   \mathsf{P} ; \tmmathbf{Q} \right)^{\dag} 
L_{i j} \left( \tmmathbf{p},
   \mathsf{P} ; \tmmathbf{Q} \right)
, \rho_i \right\} \right] .  \label{eq:rirate}
\end{eqnarray}
Here
\begin{eqnarray}
  \mathsf{H}^i_n & = & \text{$- 2 \pi \hbar^2 \frac{n_{\tmop{gas}}}{m_{\ast}}
  \bignone \int d\tmmathbf{p}_0 \mu (\tmmathbf{p}_0)$} \text{Re}  \left[ f_{i
  i} \left( \text{rel} \left( \tmmathbf{p}_0, \mathsf{P} \right), \text{rel}
  \left( \tmmathbf{p}_0, \mathsf{P} \right) \right) \right],  \label{eq:hni}
\end{eqnarray}
and $\rho_i$ denotes a collection of trace class operators in $L^2
(\mathbbm{R}^3)$ normalized according to $\sum_{i = 1}^n \tmop{Tr}_{L^2
  (\mathbbm{R}^3)} \text{$\rho_i$}=1$.


For the case in which all the off-diagonal elements with respect to momentum and internal energy
eigenvalues vanish, that is to say $\langle \tmmathbf{P}, i| \rho
|\tmmathbf{P}', k \rangle = 0$ if $\tmmathbf{P} \neq \tmmathbf{P}'$
{\tmem{{\tmem{or}}}} $i \neq k$, the motional state of the test particle is
fully characterized by the distribution of the diagonal terms $f_i (\tmmathbf{P})= \langle \tmmathbf{P}, i| \rho |\tmmathbf{P}, i
  \rangle$, which is a classical probability density obeying the
classical Markovian master equation
\begin{eqnarray}
  \frac{\mathd}{\tmop{dt}} f_i (\tmmathbf{P}) & = & \sum_j \int
  d\tmmathbf{Q}f_j (\tmmathbf{P}-\tmmathbf{Q}) M^{j j}_{i i} (\tmmathbf{P};
  \tmmathbf{Q}) - f_i (\tmmathbf{P}) \sum_j \int d\tmmathbf{Q}M^{i i}_{j j}
  (\tmmathbf{P}+\tmmathbf{Q}; \tmmathbf{Q}),  \label{eq:classicaleq}
\end{eqnarray}
where the positive quantities $M^{j j}_{i i} (\tmmathbf{P};
\tmmathbf{Q})$ defined in Eq.~(\ref{eq:clarate}) can actually be
interpreted as the transition rates from an initial momentum
$\tmmathbf{P}-\tmmathbf{Q}$ and internal state $j$ to a final momentum
$\tmmathbf{P}$ and internal state $i$.  This classical Markovian
master equation provides the natural generalization of the classical
linear Boltzmann equation to a particle with internal degrees of
freedom {\cite{Snider1998a}}.

\section{Effect of internal degrees of freedom on centre of mass
decoherence\label{sec:decoh}}

The quantum linear Boltzmann equation has proven useful in the description of
collisional decoherence, as well as in the evaluation of the index of
refraction for matter waves
{\cite{Hornberger2003a,Champenois2008a,Jacquey2007a,Hornberger2008a,Vacchini2009a}}.
We will now consider the effect of internal degrees of freedom, affecting the
collisional interaction between massive test particle and background gas, on
the visibility of the interference fringes in a interferometric setup. In
particular we will show that the visibility can exhibit oscillations due to
non-Markovian effects. The effect of the entanglement between internal and
centre of mass degrees of freedom for the visibility of quantum interference
experiments has already been considered in {\cite{Hillary2005a}},  in the
absence however of decoherence effects.

\subsection{Generalized Lindblad structure\label{sec:gls}}

The quantum master equation Eq.~(\ref{eq:qlbefinal}) is in Lindblad form: this
means that the dynamics of the test particle is Markovian when both
translational and internal degrees of freedom are described and detected. A
different situation emerges if the translational or the internal degrees of
freedom, although influencing the collisional dynamics, are not revealed
during the measurement process. In this case they must be averaged out from
the description of the system, by means of the partial trace, thus becoming
part of the environment. As well known in the classical case, a non-Markovian
dynamical regime becomes Markovian by suitably enlarging the set of degrees of
freedom and vice-versa. Indeed, a unitary Markovian time evolution for both
system and reservoir generally gives a non-Markovian reduced dynamics for the
system, the degree of non-Markovianity of the description also depending on
where we set the border between system and environment, which ultimately
depends on the physical quantities actually measurable by the experimenter. A
smaller set of observed degrees of freedom, with respect to those actually
involved in the dynamics, can lead from a Markovian to a non-Markovian regime.
A general mechanism describing this passage in quantum systems is presented in
{\cite{Vacchini2008a}}: a Lindblad structure on a bipartite system can
generate in the two reduced subsystems a generalized Lindblad structure,
typically describing a non-Markovian dynamics.

In the situation we are considering, the bipartite system is formed by the
translational and the internal degrees of freedom of the test particle. If the
measurements at the output of the detector cannot probe the internal degrees
of freedom, the only experimentally accessible quantities are expectations or
matrix elements of the statistical operator given by
\begin{eqnarray}
  \varrho (t) & = \text{Tr}_{\mathbbm{C}^n} \left\{ \rho (t) \right\} = \sum_i
  \langle i| \rho (t) |i \rangle = : & \sum_i \rho_i (t),  \label{eq:varrho}
\end{eqnarray}
where $\rho (t)$ is the statistical operator describing the full dynamics of
the test particle. It is easy to see that, if the free Hamiltonian is non
degenerate, the diagonal matrix elements in the energy basis with respect to
the internal degrees of freedom of the master equation lead to
Eq.~(\ref{eq:rirate}), i.e. a coupled system of equations for the collection
$\left\{ \rho_i (t) \right\}_i$ of trace class operators on $L^2
(\mathbbm{R}^3)$. This system of equations has a generalized Lindblad
structure {\cite{Breuer2007a}}, and therefore it can also describe highly
non-Markovian dynamics for the statistical operator $\varrho (t)$ given by
Eq.~(\ref{eq:varrho}). Indeed, there is generally no closed evolution equation
for $\varrho (t)$, but from the knowledge of the initial collection $\left\{
\rho_i (0) \right\}_i$ the generalized Lindblad structure allows to obtain the
collection $\left\{ \rho_i (t) \right\}_i$ at time $t$ and therefore, through
Eq.~(\ref{eq:varrho}), also $\varrho (t)$. In the next paragraph we are going
to explicitly point out non-Markovian behaviour described by the generalized
Lindblad structure, which express the effect of correlations between internal
and translational degrees of freedom on the visibility of interference fringes
for superpositions of motional states. A complementary situation has been
considered in {\cite{Hemming2010a}}, where the effect of collisional
decoherence on internal state superpositions of a cold gas has been studied in
detail.

In typical interferometric experiments the test particle is much more massive
than the particles of the background gas. The dependence on the momentum
operator in the Lindblad operators describing the collisional dynamics and in the
Hamiltonian part determining the energy shift can therefore be replaced by a
fixed value $\tmmathbf{P}_0$, which represents the initial momentum of the
test particle entering the interferometer. 
Taking the diagonal matrix elements of the general form of the master
equation given by Eq.~(\ref{eq:qlbefinal}) and specified by
Eq.~(\ref{eq:hn})  and Eq.~(\ref{eq:diag}), assuming non degeneracy of
the internal energy eigenvalues one finally obtains for the collection
of operators $\rho_i (t)=\langle i| \rho (t) |i \rangle$ the following
coupled system of equations:
\begin{eqnarray}
  \frac{\mathd}{\tmop{dt}} \rho_i (t) & = & \frac{1}{i \hbar} \left[
  \frac{\mathsf{P}^2}{2 M}, \rho_i (t) \right] + \sum_j \left( \Gamma^{i
  j}_{\tmmathbf{P}_0} \int d\tmmathbf{Q} \mathcal{P}^{i j}_{\tmmathbf{P}_0}
  (\tmmathbf{Q}) \mathe^{i\tmmathbf{Q} \cdot \mathsf{X} / \hbar}  \rho_j (t)
  \mathe^{-i\tmmathbf{Q} \cdot \mathsf{X} / \hbar} - \Gamma^{j
  i}_{\tmmathbf{P}_0} \rho_i (t) \right),   \label{eq:lgen}
\end{eqnarray}
with $\mathcal{P}^{i j}_{\tmmathbf{P}}$ and $\Gamma^{j i}_{\tmmathbf{P}}$
probability densities and transition rates defined by
\begin{eqnarray}
  \mathcal{P}^{i j}_{\tmmathbf{P}} (\tmmathbf{Q}) & : = & \frac{M^{j j}_{i i}
  (\tmmathbf{P}; \tmmathbf{Q})}{\int d\tmmathbf{Q}M^{j j}_{i i}
  (\tmmathbf{P}+\tmmathbf{Q}; \tmmathbf{Q})}  \label{eq:pij}
\end{eqnarray}
and
\begin{eqnarray}
  \Gamma^{i j}_{\tmmathbf{P}} & : = & \int d\tmmathbf{Q}M^{j j}_{i i}
  (\tmmathbf{P}+\tmmathbf{Q}; \tmmathbf{Q}) .  \label{eq:jij}
\end{eqnarray}
respectively. 
Note that at variance with Eq.~(\ref{eq:rirate}) we are now not
assuming a classical dynamics for the internal degrees of freedom, we
focus on the diagonal matrix elements of the internal states only
since the latter are enough to determine the non-Markovian dynamics of
the motional state according to Eq.~(\ref{eq:varrho}).
The fact that the positive quantities $M^{j j}_{i
i} (\tmmathbf{P}; \tmmathbf{Q})$ are transition rates implies that $\mathcal{P}^{i
j}_{\tmmathbf{P}} (\tmmathbf{Q})$ can be interpreted as the probability
distribution function for a test particle with momentum $\tmmathbf{P}$ and
internal energy eigenstate $j$ to exchange a momentum $\tmmathbf{Q}$, and to
go into the internal state $i$ due to a collision with the gas. On the same
footing, $\Gamma^{i j}_{\tmmathbf{P}}$ can be interpreted as the total
transition rate for a test particle with momentum $\tmmathbf{P}$ and internal
state $j$ to go to a fixed final internal energy eigenstate $i$.

\subsection{Explicit solutions in position representation}
\label{explicit}

We are now going to describe the visibility reduction predicted by the
generalized Lindblad structure Eq.~(\ref{eq:lgen}) obtained from the quantum
Bloch-Boltzmann equation Eq.~(\ref{eq:qlbefinal}) in the limit of a very
massive test particle.

To obtain the formula describing the fringes visibility in an explicit way we
need to solve the equation of motion in the position representation. Starting
from Eq.~(\ref{eq:lgen}) and omitting for simplicity the explicit dependence
on the classical label $\tmmathbf{P}_0$ denoting the momentum of the test
particle, we obtain
\begin{eqnarray}
  \frac{\mathd}{\tmop{dt}} \rho_i (\tmmathbf{X}, \tmmathbf{X}', t)  & = &
  \frac{1}{i \hbar} (\Delta_{\tmmathbf{X}} - \Delta_{\tmmathbf{X}'}) \rho_i
  (\tmmathbf{X}, \tmmathbf{X}', t) \nonumber\\
  &  & + \sum_j \left( \Gamma^{i j} \Phi^{i j} \left(
  \tmmathbf{X}-\tmmathbf{X}' \right) \rho_j (\tmmathbf{X}, \tmmathbf{X}', t) -
  \Gamma^{j i} \rho_i (\tmmathbf{X}, \tmmathbf{X}', t) \right),
  \label{eq:rhoixx}
\end{eqnarray}
where $\rho_i (\tmmathbf{X}, \tmmathbf{X}', t)$ denotes the matrix element
$\langle \tmmathbf{X}| \rho_i (t) |\tmmathbf{X}' \rangle$ and $\Phi^{i j}
\left( \tmmathbf{X}-\tmmathbf{X}' \right)$ is the characteristic function of
the probability density $\mathcal{P}^{i j} (\tmmathbf{Q})$
{\cite{Feller1971}}, i.e. its Fourier transform
\begin{eqnarray}
  \Phi^{i j} \left( \tmmathbf{X}-\tmmathbf{X}' \right) & = & \int
  d\tmmathbf{Q}e^{i (\tmmathbf{X}-\tmmathbf{X}') \cdot \tmmathbf{Q}/ \hbar}
  \mathcal{P}^{i j} (\tmmathbf{Q}) .  \label{eq:fi}
\end{eqnarray}
We will now consider a few cases in which Eq.~(\ref{eq:rhoixx}) can be solved
analytically, so as to obtain an exact expression for the visibility, showing
up different possible qualitative behaviour.

\subsubsection{N-level system}

When the collisions are purely elastic, so that they do not lead to
transitions between different internal states, the scattering rates satisfy
$\Gamma^{i j} = \delta_{i j} \Gamma^{i i}$. This is the case when the energy
exchanges involved in the single collisions are much smaller than the typical
separation of the internal energy levels {\cite{Alicki2003a}}. The equations
for the different $\rho_i$ then become uncoupled and take the form
\begin{eqnarray}
  \frac{\mathd}{\tmop{dt}} \rho_i (\tmmathbf{X}, \tmmathbf{X}', t) & = &
  \frac{1}{i \hbar} (\Delta_{\tmmathbf{X}} - \Delta_{\tmmathbf{X}'}) \rho_i
  (\tmmathbf{X}, \tmmathbf{X}', t) - \Gamma^{i i} (1 - \Phi^{i i} \left(
  \tmmathbf{X}-\tmmathbf{X}' \right)) \rho_i (\tmmathbf{X}, \tmmathbf{X}', t)
  .  \label{eq:decouple}
\end{eqnarray}
The latter equation can be conveniently solved introducing the function
{\cite{Savage1985a}}
\begin{eqnarray}
  \chi_i (\tmmathbf{\lambda}, \tmmathbf{\mu}, t) & : = & \text{Tr} \left\{
  \rho_i (t) e^{i (\tmmathbf{\lambda} \cdot \mathsf{X} +\tmmathbf{\mu} \cdot
  \mathsf{P}) / \hbar} \right\}  \label{eq:chi}
\end{eqnarray}
where $\mathsf{X}$ and $\mathsf{P}$ as usual denote position and momentum
operators of the test particle. In such a way Eq.~(\ref{eq:decouple}) leads to
\begin{eqnarray}
  \partial_t \chi_i (\tmmathbf{\lambda}, \tmmathbf{\mu}, t) & = & \left[
  \frac{\tmmathbf{\lambda}}{M} \cdot \partial_{\tmmathbf{\mu}} - \Gamma^{i i}
  (1 - \Phi^{i i} (\tmmathbf{\mu})) \right] \chi_i (\tmmathbf{\lambda},
  \tmmathbf{\mu}, t),
\end{eqnarray}
which is an equation of first order solved by
\begin{eqnarray}
  \chi_i (\tmmathbf{\lambda}, \tmmathbf{\mu}, t) & = & \chi^0_i
  (\tmmathbf{\lambda}, \tmmathbf{\lambda}t / M +\tmmathbf{\mu}) e^{- \Gamma^{i
  i} \int^t_0 (1 - \Phi^{i i} (\tmmathbf{\lambda}(t - t') / M
  +\tmmathbf{\mu})) \tmop{dt}'},  \label{eq:chiusa}
\end{eqnarray}
where the function $\chi_i^0 (\tmmathbf{\lambda}, \tmmathbf{\lambda}t / M
+\tmmathbf{\mu})$ obeys the free equation $\partial_t \chi_i (\tmmathbf{\lambda}, \tmmathbf{\mu}, t) = 
  ({\tmmathbf{\lambda}}/{M}) \cdot \partial_{\tmmathbf{\mu}}  \chi_i
  (\tmmathbf{\lambda}, \tmmathbf{\mu}, t)$.
Inverting Eq.~(\ref{eq:chi}) by taking the Fourier transform with respect to
$\tmmathbf{\lambda}$,
\begin{eqnarray}
  \rho_i (\tmmathbf{X}, \tmmathbf{X}', t) & = & \int
  \frac{d\tmmathbf{\lambda}}{(2 \pi \hbar)^3} e^{- i\tmmathbf{\lambda} \cdot
  (\tmmathbf{X}+\tmmathbf{X}') / 2 \hbar} \chi_i (t, \tmmathbf{\lambda},
  \tmmathbf{X}-\tmmathbf{X}'),  \label{eq:fourier}
\end{eqnarray}
we obtain the exact solution
\begin{eqnarray*}
  \rho_i (\tmmathbf{X}, \tmmathbf{X}', t) & = & \int
  \frac{d\tmmathbf{s}d\tmmathbf{\lambda}}{(2 \pi \hbar)^3} e^{-
  i\tmmathbf{\lambda} \cdot \tmmathbf{s}/ \hbar} e^{- \Gamma^{i i} \int^t_0 (1
  - \Phi^{i i} ( \tmmathbf{\lambda}(t - t') / M +\tmmathbf{X}-\tmmathbf{X}'))
  \tmop{dt}'} \rho_i^0 (\tmmathbf{X}+\tmmathbf{s}, \tmmathbf{X}'
  +\tmmathbf{s}, t)
\end{eqnarray*}
expressed in terms of an integral of the freely evolved subcollections
$\rho_i^0 (\tmmathbf{X}, \tmmathbf{X}', t)$ with a suitable kernel, where we
have set
\begin{eqnarray}
  \rho_i^0 (\tmmathbf{X}, \tmmathbf{X}', t) & = & \int
  \frac{d\tmmathbf{\lambda}}{(2 \pi \hbar)^3} e^{- i\tmmathbf{\lambda} \cdot
  (\tmmathbf{X}+\tmmathbf{X}') / 2 \hbar} \chi^0 (\tmmathbf{\lambda},
  \tmmathbf{X}-\tmmathbf{X}', t)  \nonumber\\
  & = & \langle \tmmathbf{X}| \exp \left( - \frac{i}{\hbar}
  \frac{\mathsf{P}^2}{2 M} t \right) \rho_i (0) \exp \left( + \frac{i}{\hbar}
  \frac{\mathsf{P}^2}{2 M} t \right) |\tmmathbf{X}' \rangle \label{eq:Sfree}
\end{eqnarray}
and $\rho_i (0) = \langle i| \rho (0) |i \rangle$. The evolution of the
statistical operator given by Eq.~(\ref{eq:varrho}) is obtained summing the
different $\rho_i (\tmmathbf{X}, \tmmathbf{X}', t)$ over the discrete index
$i$. For an initial state given by a product state between the translational
and the internal part, so that $\rho_i (0) = p_i \varrho (0)$, we finally
obtain
\begin{eqnarray}
  \varrho (\tmmathbf{X}, \tmmathbf{X}', t) & = & \sum_i p_i \int
  \frac{d\tmmathbf{s}d\tmmathbf{\lambda}}{(2 \pi \hbar)^3} e^{-
  i\tmmathbf{\lambda} \cdot \tmmathbf{s}/ \hbar} e^{- \Gamma^{i i} \int^t_0 (1
  - \Phi^{i i} (\tmmathbf{\lambda}(t - t') / M +\tmmathbf{X}-\tmmathbf{X}'))
  \tmop{dt}'} \varrho^0 (\tmmathbf{X}+\tmmathbf{s}, \tmmathbf{X}'
  +\tmmathbf{s}, t) .\nonumber\\  \label{eq:varrhoxx}
\end{eqnarray}
This result reduces to the standard Markovian situation, when either only one
$p_i$ is different from zero (and therefore equal to one), or the rates are
all equal. This limiting cases describes situations in which the initial
state is in a specific internal state or the collisions do not depend on the
internal state of the tracer particle.

\subsubsection{Two-level system}
\label{2ls}

For the case of a two-level system a natural situation corresponds to
inelastic scattering taking place only when the test particle gets de-excited,
so that only one of the two scattering rates is different from zero. This case
can still be treated analytically. Assuming $\Gamma^{21} = 0$, the equation
for $\chi_2 (t, \tmmathbf{\lambda}, \tmmathbf{\mu})$ gets closed, and is
solved by
\begin{eqnarray}
  \chi_2 (\tmmathbf{\lambda}, \tmmathbf{\mu}, t) & = & \chi^0_2
  (\tmmathbf{\lambda}, \tmmathbf{\lambda}t / M +\tmmathbf{\mu}) e^{- \Gamma^{1
  2}_{} t} e^{- \Gamma^{2 2} \int^t_0 (1 - \Phi^{2 2} (\tmmathbf{\lambda}(t -
  t') / M +\tmmathbf{\mu})) \tmop{dt}'} .  \label{eq:chi1}
\end{eqnarray}
The equation for $\chi_1 (\tmmathbf{\lambda}, \tmmathbf{\mu}, t)$ then reads
\begin{eqnarray}
  \partial_t \chi_1 (\tmmathbf{\lambda}, \tmmathbf{\mu}, t) & = & \left[
  \frac{\tmmathbf{\lambda}}{M} \cdot \partial_{\tmmathbf{\mu}} -
  \Gamma^{11}_{} (1 - \Phi^{11} (\tmmathbf{\mu})) \right] \chi_1
  (\tmmathbf{\lambda}, \tmmathbf{\mu}, t) + \Gamma^{12} \Phi^{12}
  (\tmmathbf{\mu}) \chi_2 (\tmmathbf{\lambda}, \tmmathbf{\mu}, t)
  \label{eq:chi2}
\end{eqnarray}
and its solution is given by
\begin{eqnarray}
  \chi_1 (\tmmathbf{\lambda}, \tmmathbf{\mu}, t) & = & e^{ -
  \Gamma^{11} \int^t_0 (1 - \Phi^{11} ( \tmmathbf{\lambda}(t - t') / M
  +\tmmathbf{\mu})) \tmop{dt}' } \left\{ \chi^0_1 (\tmmathbf{\lambda},
  \tmmathbf{\lambda}t / M +\tmmathbf{\mu}) \right. \nonumber\\
  &  & + \Gamma_{}^{12} \int^t_0 \left[ e^{ \Gamma^{1 1} \int^{t'}_0
  (1 - \Phi^{11} ( \tmmathbf{\lambda}(t - t'') / M +\tmmathbf{\mu}))
  \tmop{dt}'' } \right. \nonumber\\
  &  & \left. \times \Phi^{12} ( \tmmathbf{\lambda}(t - t') / M
  +\tmmathbf{\mu}) \chi_2 (t', \tmmathbf{\lambda}, \tmmathbf{\lambda}(t - t')
  / M +\tmmathbf{\mu})] d t' \phantom{\frac{1}{1}} \right\} .  \label{eq:chi3}
\end{eqnarray}
This formula explicitly shows that $\chi_1 (\tmmathbf{\lambda},
\tmmathbf{\mu}, t)$ depends on the function $\chi_2 (\tmmathbf{\lambda},
\tmmathbf{\mu}, \cdot)$ evaluated over the whole time interval between 0 and
$t$, a typical signature of non-Markovian dynamics. Assuming once again that
the initial state is characterized by $\rho_i (0) = p_i \varrho_{} (0)$, the
statistical operator describing the translational degrees of freedom of the
test particle is given at time $t$ by the expression
\begin{eqnarray}
  \varrho (\tmmathbf{X}, \tmmathbf{X}', t) & = & \int
  \frac{d\tmmathbf{s}d\tmmathbf{\lambda}}{(2 \pi \hbar)^3} e^{-
  i\tmmathbf{\lambda} \cdot \tmmathbf{s}/ \hbar} \varrho_{}^0
  (\tmmathbf{X}+\tmmathbf{s}, \tmmathbf{X}' + s, t) \left\{ p_2 e^{-
  \Gamma^{12} t} e^{- \Gamma^{22} \int^t_0 (1 - \Phi^{22}
  (\tmmathbf{\lambda}(t - t') / M +\tmmathbf{X}-\tmmathbf{X}')) \tmop{dt}'}
  \right. \nonumber\\
  &  & + p_1 e^{- \Gamma^{11} \int^t_0 (1 - \Phi^{11} (\tmmathbf{\lambda}(t -
  t') / M +\tmmathbf{X}-\tmmathbf{X}')) \tmop{dt}'}+ p_2 \Gamma^{12} e^{- \Gamma^{11} \int^t_0 (1 - \Phi^{11}
  (\tmmathbf{\lambda}(t - t') / M +\tmmathbf{X}-\tmmathbf{X}')) \tmop{dt}'}
  \nonumber\\
  &  & \times \int^t_0 \left( e^{- \Gamma^{1 2} t'} e^{- \Gamma^{22}
  \int^{t'}_0 (1 - \Phi^{22} (\tmmathbf{\lambda}(t - t'') / M
  +\tmmathbf{X}-\tmmathbf{X}')) \tmop{dt}''} e^{ \Gamma^{11} \int^{t'}_0 (1 -
  \Phi^{11} (\tmmathbf{\lambda}(t - t'') / M +\tmmathbf{X}-\tmmathbf{X}'))
  \tmop{dt}''} \right. \nonumber\\
  &  & \left. \left. \times \Phi^{12} (\tmmathbf{\lambda}(t - t') / M
  +\tmmathbf{X}-\tmmathbf{X}') \phantom{\frac{1}{}} \right) d t' \right\} .
  \label{eq:varrhoxx3}
\end{eqnarray}

\subsection{Non-exponential visibility reduction }

We can now explicitly present the visibility reduction predicted by
the generalized Lindblad structure obtained from the quantum master
equation for a test particle with internal degrees of freedom.  Our
aim is to obtain an exact expression for the loss of visibility in a
double-slit arrangement as a function of the time of interaction with
the environment, and to illustrate by means of example how the
presence of the various scattering channels, corresponding to the
different internal states, can actually lead to non-Markovian
behaviours.  In particular we will consider the situation of purely
elastic collisions in full generality, also allowing for inelastic
scattering in the case of a two-level system. While the experimental
setting is always taken to be the same, the different number of
internal degrees of freedom involved and the presence or absence of
inelastic scattering events will lead to more or less marked
non-exponential behaviours in the reduction of the visibility fringes.

\subsubsection{Visibility formula}

We first derive a formula for the visibility reduction in the case of
a double-slit experiment in the far field approximation. A beam of
particles moves towards a grating perpendicular to its direction of
propagation, and with two identical slits separated by a distance $d$,
finally reaching a detector where the fringes of interference are
observed.  During the flight through the interferometer the beam
particles interact through collisions with the environment in the
background, thus undergoing decoherence.  We consider an initial
product state, so that in the notation of Eq.~(\ref{eq:lgen}) one has
$\rho_i (0) = p_i \varrho (0)$, where $\varrho (0)$ describes the
translational degrees of freedom. If after the passage through the
collimation slits the test particle is described by
$\rho_{\tmop{sl}}$, then the double-slit grating prepares the initial
state {\cite{Hornberger2006a}}
\begin{eqnarray}
  \varrho (0) & = & 2 \cos \left( \frac{\mathsf{P} \cdot \tmmathbf{d}}{2
  \hbar} \right) \rho_{\tmop{sl}} \cos \left( \frac{\mathsf{\mathsf{P}} \cdot
  \tmmathbf{d}}{2 \hbar} \right) .  \label{eq:cos}
\end{eqnarray}
Setting
\begin{eqnarray}
  \varrho (\tmmathbf{X}, \tmmathbf{X}, t) & = & \langle
  \tmmathbf{X}|\mathcal{U}(t) [\varrho (0)] |\tmmathbf{X} \rangle \assign I
  \left( \tmmathbf{X} \right) \nonumber
\end{eqnarray}
we consider the quantity
\begin{eqnarray*}
  \mathcal{V} & = & \frac{I_{\max} - I_{\min}}{I_{\max} + I_{\min}},
\end{eqnarray*}
which describes the reduction of the interference pattern with respect to the
free case. Exploiting the fact that the time evolution generated by
Eq.~(\ref{eq:varrhoxx}) is covariant under translations
{\cite{Vacchini2009a,Vacchini2010a}}, so that
\begin{eqnarray}
  \mathcal{U}(t) [e^{i \mathsf{P} \cdot \tmmathbf{a}/ \hbar} \varrho e^{- i
  \mathsf{P} \cdot \tmmathbf{a}/ \hbar}] & = & e^{i \mathsf{P} \cdot
  \tmmathbf{a}/ \hbar} \mathcal{U}(t) [\varrho] e^{- i \mathsf{P} \cdot
  \tmmathbf{a}/ \hbar},  \label{eq:cov}
\end{eqnarray}
one has, using Eq.~(\ref{eq:cos}),
\begin{eqnarray}
  \mathcal{V} & = & \frac{2 \left| \langle \tmmathbf{X}- \frac{1}{2}
  \tmmathbf{d}|\mathcal{U}(t) [\rho_{\tmop{sl}} e^{- i \mathsf{P} \mathsf{}
  \cdot \tmmathbf{d}/ \hbar}] |\tmmathbf{X}- \frac{1}{2} \tmmathbf{d} \rangle
  | \right.}{\langle \tmmathbf{X}- \frac{1}{2} \tmmathbf{d}|\mathcal{U}(t)
  [\rho_{\tmop{sl}}] |\tmmathbf{X}- \frac{1}{2} \tmmathbf{d} \rangle + \langle
  \tmmathbf{X}+ \frac{1}{2} \tmmathbf{d}|\mathcal{U}(t) [\rho_{\tmop{sl}}]
  |\tmmathbf{X}+ \frac{1}{2} \tmmathbf{d} \rangle},  \label{eq:v}
\end{eqnarray}
where now $t$ is the time employed by the test particle to reach the detector.
Indeed this result remains true for any translation-covariant time evolution.

For an initial factorized state of the test particle we can exploit
Eq.~(\ref{eq:varrhoxx}) to obtain a closed formula for the time evolution
operator $\mathcal{U}(t)$ depending on the initial internal state, i.e. on the
coefficients $p_i$ appearing in $\rho_i (0) = p_i \varrho (0)$: the numerator
of Eq.~(\ref{eq:v}) then reads
\begin{eqnarray*}
  & 2 \left| \sum_i p_i \int \frac{d\tmmathbf{s}d\tmmathbf{\lambda}}{(2 \pi
  \hbar)^3} e^{- i\tmmathbf{\lambda} \cdot \tmmathbf{s}/ \hbar} \langle
  \tmmathbf{X}- \frac{\tmmathbf{d}}{2} +\tmmathbf{s}|\mathcal{U}_0 (t)
  [\rho_{\tmop{sl}} e^{- i \mathsf{P} \cdot \tmmathbf{d}/ \hbar}]
  |\tmmathbf{X}- \frac{\tmmathbf{d}}{2} +\tmmathbf{s} \rangle e^{- \Gamma^{i
  i} \int^t_0 (1 - \Phi^{i i} (\tmmathbf{\lambda}(t - t') / M)) \tmop{dt}'} |,
  \right. &
\end{eqnarray*}
where $\mathcal{U}_0 (t)$ is the free evolution operator of the translational
degrees of freedom, so that
\begin{eqnarray}
  \langle \tmmathbf{X}- \frac{\tmmathbf{d}}{2} +\tmmathbf{s}|\mathcal{U}_0 (t)
  [\rho_{\tmop{sl}} e^{- i \mathsf{P} \cdot \tmmathbf{d}/ \hbar}]
  |\tmmathbf{X}- \frac{\tmmathbf{d}}{2} +\tmmathbf{s} \rangle & = &
  \text{$\langle \tmmathbf{X}- \frac{\tmmathbf{d}}{2} +\tmmathbf{s}|
  \mathcal{U}_0 (t) [\rho_{\tmop{sl}}] |\tmmathbf{X}+ \frac{\tmmathbf{d}}{2}
  +\tmmathbf{s} \rangle$} .  \label{eq:xmenod}
\end{eqnarray}
The latter expression can also be written
\begin{eqnarray}
  \langle \tmmathbf{X}- \frac{\tmmathbf{d}}{2} | \mathcal{U}_0 (t)
  [\rho_{\tmop{sl}}] |\tmmathbf{X}+ \frac{\tmmathbf{d}}{2} \rangle & = &
  \left( \frac{M}{t} \right)^3 e^{- iM\tmmathbf{d} \cdot \tmmathbf{X}/ \left(
  \hbar t \right)} \int \frac{d\tmmathbf{Y}d\tmmathbf{Y}'}{(2 \pi \hbar)^3}
  e^{iM \left( \tmmathbf{Y}^2 -\tmmathbf{Y}'^2 \right) / \left( 2 \hbar t
  \right)} \nonumber\\
  &  &\times e^{- iM\tmmathbf{X} \cdot (\tmmathbf{Y}-\tmmathbf{Y}') / \left( \hbar
  t \right) } e^{iM\tmmathbf{d} \cdot \left( \tmmathbf{Y}+\tmmathbf{Y}'
  \right) / \left( 2 \hbar t \right)} \langle \tmmathbf{Y}| \rho_{\tmop{sl}}
  |\tmmathbf{Y}' \rangle,  \label{eq:fraun}
\end{eqnarray}
assuming due to symmetry $\tmop{Tr} \left( \mathsf{X} \rho_{\tmop{sl}} \right)
= 0$.

This formula enables us to implement the far field approximation. In fact,
let $\sigma \tmop{be} \tmop{the} \tmop{width}$ of the two slits, so that the
integrand is negligible if $\tmmathbf{Y}$ (and similarly for $\tmmathbf{Y}'$)
takes values outside the support of $\rho_{\tmop{sl}}$, then $M\tmmathbf{Y}^2
/ \left( \hbar t \right) \lesssim M \sigma^2 / \left( \hbar t \right)$ and
therefore for a time long enough such that $ \hbar t / M \gg \sigma^2$ the
first exponential can be disregarded. The same applies for the last
exponential if $\hbar t / M \gg \sigma d$. For times longer than $\max \left\{
M \sigma^2 / \hbar, M \sigma d / \hbar \right\}$, corresponding to the far
field approximation, we get
\begin{eqnarray}
  \langle \tmmathbf{X}- \frac{\tmmathbf{d}}{2} | \mathcal{U}_0 (t)
  [\rho_{\tmop{sl}}] |\tmmathbf{X}+ \frac{\tmmathbf{d}}{2} \rangle & \approx &
  \left( \frac{M}{t} \right)^3 e^{- iM\tmmathbf{d} \cdot \tmmathbf{X}/ \left(
  \hbar t \right)}  \tilde{\rho}_{\tmop{sl}} \left( \frac{M}{t} \tmmathbf{X}
  \right),  \label{eq:approx}
\end{eqnarray}
where $\tilde{\rho}_{\tmop{sl}} \left( \cdot \right)$ is the distribution
function for the momentum of the particle in the state $\rho_{\tmop{sl}}$,
\begin{eqnarray*}
  \tilde{\rho}_{\tmop{sl}} \left( \frac{M}{t} \tmmathbf{X} \right) & = & \int
  \frac{d\tmmathbf{Y}d\tmmathbf{Y}'}{(2 \pi \hbar)^3} e^{- iM\tmmathbf{X}
  \cdot (\tmmathbf{Y}-\tmmathbf{Y}') / \left( \hbar t \right) } \langle
  \tmmathbf{Y}| \rho_{\tmop{sl}} |\tmmathbf{Y}' \rangle .
\end{eqnarray*}
The equivalence between the assumption $\hbar t / M \gg \sigma^2$ and the far
field approximation $L \gg \sigma^2 / \lambda$, where \ $\lambda = \hbar /
P_z$ is the wavelength associated to the test particle and $L$ is the distance
between grating and detector, is easily seen from the relation $L = p_z t /
M$, where $p_z$ is the component along the $z$ direction of the massive
particle, assumed to be constant. Substituting Eq.~(\ref{eq:xmenod}) in the
numerator of Eq.~(\ref{eq:v}) and using the approximation
$\tilde{\rho}_{\tmop{sl}} \left( M (\tmmathbf{X}+\tmmathbf{s}) / t \right)
\approx \tilde{\rho}_{\tmop{sl}} \left( M\tmmathbf{X}/ t \right)$ valid
because of the localization of the state $\rho_{\tmop{sl}}$, we can easily
perform the integrals over $\tmmathbf{s}$ and $\tmmathbf{\lambda}$, thus
finally obtaining
\begin{eqnarray*}
  & 2 \left( \frac{M}{t} \right)^3 |  \tilde{\rho}_{\tmop{sl}} \left(
  \frac{M}{t} \tmmathbf{X} \right) | | \sum_{i = 1}^n p_i e^{- \Gamma^{i i}
  \int^t_0 \left( 1 - \Phi^{i i} \left( \tmmathbf{d} \frac{t' - t}{t} \right)
  \right) \tmop{dt}'} | . &
\end{eqnarray*}
For the denominator of Eq.~(\ref{eq:v}) one can proceed in an analogous way,
using
\begin{eqnarray*}
  \langle \tmmathbf{X} \pm \frac{1}{2} \tmmathbf{d}+\tmmathbf{s}|\mathcal{U}_0
  (t) [\rho_{\tmop{sl}}] |\tmmathbf{X} \pm \frac{1}{2}
  \tmmathbf{d}+\tmmathbf{s} \rangle & \approx & \left( \frac{M}{t} \right)^3
  \tilde{\rho}_{\tmop{sl}} \left( \frac{M}{t} \tmmathbf{X} \right),
\end{eqnarray*}
and performing the integral over $\tmmathbf{\lambda}$, further observing that
$\Phi^{i j} \left( 0 \right) = 1$ for the normalization of $\mathcal{P}^{i j}
(\tmmathbf{Q})$. 

\subsubsection{Non-exponential behaviours}

The desired expression for the visibility in the absence of inelastic
scattering and for an arbitrary number $n$ of channels thus reads
\begin{eqnarray}
  \mathcal{V} & = & | \sum^n_{i = 1} p_i e^{- \Gamma^{i i} \int^t_0 \left( 1 -
  \Phi^{i i} \left( \tmmathbf{d} \frac{t' - t}{t} \right) \right) \tmop{dt}'}
  | ,  \label{eq:vv}
\end{eqnarray}
where we recall that the probabilities $p_i$ give the weight of the
different internal states in the initial preparation.
The dependence on $t$ in this formula can be easily made explicit with the
change of variable $t' / t = s$, so that one has
\begin{eqnarray}
  \mathcal{V} & = & | \sum^n_{i = 1} p_i e^{- \Gamma^{i i}  \left( 1 -
  \int^1_0 \Phi^{i i} \left( \tmmathbf{d}(s - 1) \right) d s \right) t }  | .
  \label{eq:vvt}
\end{eqnarray}

\begin{figure}[ht]
  \includegraphics[scale=0.91]{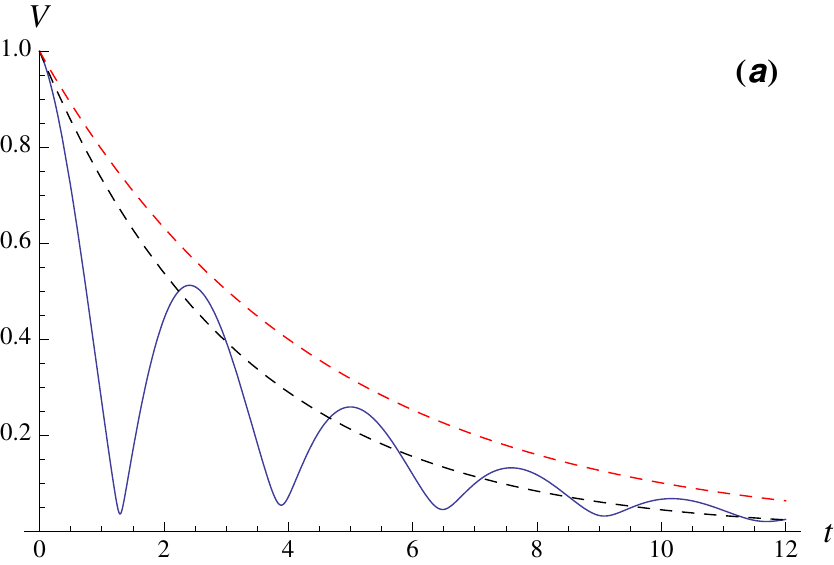} \includegraphics[scale=0.91]{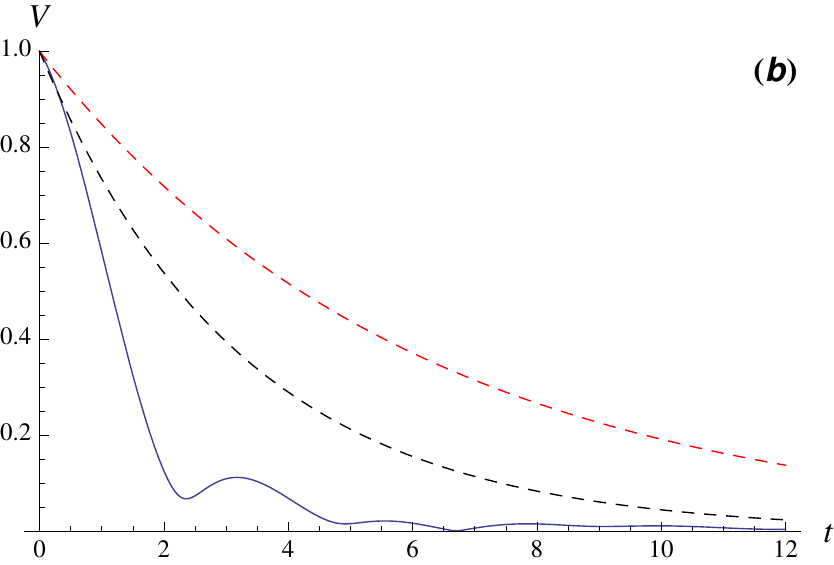}
  \caption{\label{fig:1} Plot of the visibility in a double-slit
    arrangement as a function of the interaction time with the
    environment, for the case of elastic scattering events only,
    according to Eq.~(\ref{eq:vvt}) and with growing number of
    channels from left to right. The dashed lines
    represent the Markovian exponential decays occurring if a single
    elastic channel prevails on the others, the one with the highest
    and lowest decay rate corresponding to lower and upper line
    respectively.  
\textbf{(a)} Visibility for $n=2$ elastic channels,
    according to the expression Eq.~(\ref{eq:v3}). It appears a non
    monotonic decay as a consequence of the
    interference between the contributions of the two different
    elastic channels. The coefficients $\alpha^i$ and $\beta^i$
    defined in Eq.~(\ref{eq:ab}) are calculated for two Gaussian
    distributions $\mathcal{P}^{1 1} (\tmmathbf{Q})$ and
    $\mathcal{P}^{2 2} (\tmmathbf{Q})$ of the exchanged momenta.
    Taking $\tmmathbf{d}= d \hat{z}$ as direction of propagation inside the
    interferometer we only need to specify the mean and the variance
    of the exchanged momenta along this axis, respectively $\mu_{i i}$
    and $\sigma_{i i}$, $i = 1, 2$. The plot is for $p_1 = p_2 =
    \frac{1}{2}$, while $\Gamma_{}^{11} = \Gamma_{}^{22} = 10$, $d =
    1$, $\sigma_{11} = \sigma_{22} = 0.1$, $\mu_{11} = - 0.2$,
    $\mu_{22} = 0.3$ in arbitrary units. 
\textbf{(b)} Visibility for $n=8$ elastic channels according to the
    general expression Eq.~(\ref{eq:vvt}). The characteristic functions $\Phi^{i i}$ are calculated
    starting from Gaussian distributions, assuming equal rates
    $\Gamma^{i i} = 10$ and equal variances $\sigma_{i i} = 0.1$ in
     arbitrary units as in (a). The $p_i$ are uniformly
    distributed and the means $\mu_{i i}$ are equally spaced in the
    range from $-0.2$ to $0.3$.}
\end{figure}

From Eq.~(\ref{eq:vvt}) one can easily see the difference between the
Markovian situation, corresponding to $n = 1$, and the
general case. If there is just one term in the sum, the modulus simply picks
out the real part of the characteristic function in the exponential and
Eq.~(\ref{eq:vvt}) describes an exponential decay in time with a rate $\Gamma
\left( 1 - \int^1_0 \tmop{Re} \left\{ \Phi \left( \tmmathbf{d}(s - 1) \right)
\right\} d s \right)$. This can happen if only one internal
energy state is populated in the initial preparation or the scattering
events are actually independent on the internal state.
If there are at least two terms, the modulus can
generate oscillating terms as a consequence of the interference of the
different phases arising since the functions $\Phi^{i i}$ are in general
complex valued. Even if the imaginary parts of the characteristic functions
are zero, i.e. the distribution functions of the exchanged momenta are even,
Eq.~(\ref{eq:vvt}) can describe highly non-exponential behaviour. In this case
in fact it reduces to
\begin{eqnarray}
  \mathcal{V} & = & \sum^n_{i = 1} p_i e^{- \Gamma^{i i} \left( 1 - \int^1_0
  \Phi^{i i} \left( \tmmathbf{d}(s - 1) \right) d s \right) t},
  \label{eq:powerlaw}
\end{eqnarray}
i.e. the sum of different exponential functions. As shown in
{\cite{Alemany1997a,Vacchini2008a}} this kind of relations can describe
behaviour very different from the exponential one.

Let us consider in more detail the case of a two-level system. Introducing
the notation
\begin{eqnarray}
  \alpha^i & \assign & \tmop{Re} \int^1_0  \left\{ \Phi^{i i} \left(
  \tmmathbf{d}(s - 1) \right) \right\} d s \nonumber \\
  \beta^i & \assign & \tmop{Im} \int^1_0 \left\{ \Phi^{i i} \left(
  \tmmathbf{d}(s - 1) \right) \right\} d s, \label{eq:ab}
\end{eqnarray}
the visibility reduction is explicitly given by
\begin{eqnarray}
  \mathcal{V} & = & \left[p_1^2 e^{- 2 \Gamma^{11} (1 - \alpha^1) t} + p_2^2
  e^{- 2 \Gamma^{22} (1 - \alpha^2) t} + 2 p_1 p_2 e^{- \Gamma^{11} (1 -
  \alpha^1) t} e^{- \Gamma^{22} (1 - \alpha^2) t} \cos \left[ ( \Gamma^{11}
  \beta^1 - \Gamma^{22} \beta^2) t \right]\right]^{1/2} . \nonumber\\ \label{eq:v3}
\end{eqnarray}
This formula describes a decrease modulated by the oscillations
produced by the cosine function. To illustrate this behaviour in
Fig.~\ref{fig:1}.a we plot the visibility as a function of time, considering by means of example two Gaussian
distributions. Note that the appearance of the oscillations depends on
a non vanishing mean value for the distribution functions
$\mathcal{P}^{i i} (\tmmathbf{Q})$ given by Eq.~(\ref{eq:pij}), which
describe the state dependent momentum transfers.  This feature
corresponds to a preferred direction in the net momentum transfer
between test particle and environment, as happens e.g. by the
interaction with a laser beam \cite{Kokorowski2001a}, the asymmetry in
the single interaction channel being determined in this case by the
direction of propagation.

The behaviour described by Eq.~(\ref{eq:vvt}) for an $n$-level system
is illustrated in Fig.~\ref{fig:1}.b, where we show how the increased
number of levels can strongly suppress the oscillations and lead to a
reduction of the visibility.  The dashed lines represent the
exponential decays pertaining to the Markovian situation arising if
only one of the internal energy states is initially populated, the one
with the highest or lowest decoherence rate corresponding to the lower
or upper dashed line respectively.  It appears that with growing $n$
the interference between the contributions of the different channels
to Eq.~(\ref{eq:vvt}) rapidly determines a decay of the visibility
sensibly faster than that occurring for the corresponding Markovian
single-channel dynamics. Indeed in Fig.~\ref{fig:1} left and right
panel correspond to the same interaction strength but differ in the number of involved
degrees of freedom, ranging to $n=2$ to $n=8$.

\begin{figure}[ht]
  \includegraphics[scale=0.95]{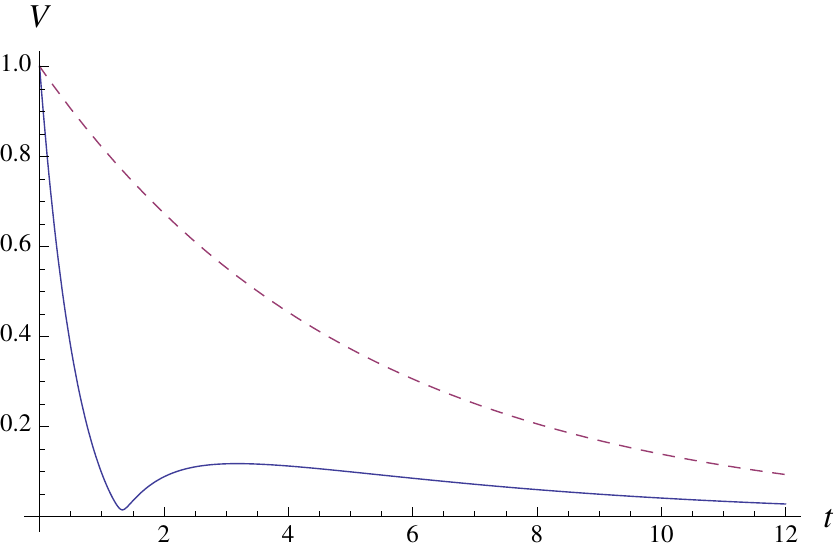}
  \caption{\label{fig:3} Plot of the visibility in a double-slit
    arrangement as a function of the interaction time with the
    environment, for the case in which one of the internal states also
    undergoes inelastic scattering, according to
    Eq.~(\ref{eq:inelastic}) with $n=2$. It clearly appears a non
    monotonic decay of the visibility as a consequence of the multiple
    time integration describing the contribution of the inelastic
    channel. The distributions of momentum transfers are assumed
    Gaussian, with $\sigma_{11}=1, \sigma_{12}=3, \mu_{11}=1,
    \mu_{12}=5$; moreover $\Gamma^{11}=0.75$ and $\Gamma^{12}=1.75$.
    The dashed line corresponds to the Markovian dynamics determined
    by the channel undergoing elastic scattering only.}
\end{figure}

Relying on the results of Sect.\ref{2ls} one can also obtain an
expression of the visibility in the presence of inelastic scattering
for a two-level system. 
Indeed starting 
from Eq.~(\ref{eq:varrhoxx3}) and following the same procedure as
above one comes to
\begin{equation}
\label{eq:inelastic}
  \mathcal{V} =  \left| 
{e^{- \Gamma^{12} t} + \Gamma^{12} e^{- \Gamma^{11} t
  \int^1_0 (1 - \Phi^{11} (\tmmathbf{d}(s - 1)) \tmop{ds}} \int^t_0 \left(
  e^{- \Gamma^{1 2} t'} e^{+ \Gamma^{11} \int^{t'}_0 (1 - \Phi^{11}
  (\tmmathbf{d}(t'' - t) / t)) \tmop{dt}''} \Phi^{12} \left( \tmmathbf{d}
  \frac{t' - t}{t} \right) \right) d t' 
}
\right| ,
\end{equation}
where for simplicity $p_2=1$, and we have taken
$\Gamma^{22} = 0$, so that the oscillations in the visibility cannot
be traced back to interference among different components.  An
illustration of the behaviour of the visibility in this case has been
plotted in Fig.~\ref{fig:3}, always assuming for the sake of
generality a Gaussian distribution of momentum transfers. In this case
the dashed line corresponds to the exponential Markovian decay
occurring if only the elastic channel is involved in the dynamics. It
immediately appears that a non monotonic behaviour in the loss of
visibility is observed also in this case, due to the multiple time
integration in Eq.~(\ref{eq:varrhoxx3}).

We have here considered the visibility reduction as a function of time.
However in typical interferometric experiments the time of flight is fixed,
and it is more natural to study the loss of visibility as a function of the
strength of the interaction with the environment. In collisional decoherence
this depends on the number of collisions, directly proportional to the gas
density or equivalently to its pressure for a fixed time of flight. As in the
Markovian case, we can thus express the visibility as a function of the
pressure of the background gas, which is the physical quantity directly
tunable in actual experiments {\cite{Hornberger2003a}}. Introducing the
effective cross section $\sigma_{\tmop{eff}} (\tmmathbf{P}_0, i)$ according to
the relation {\cite{Vacchini2009a}}
\begin{eqnarray}
  \sum_j \int d\tmmathbf{Q}M^{i i}_{j j} (\tmmathbf{P}+\tmmathbf{Q};
  \tmmathbf{Q}) & = & n_{\tmop{gas}} \frac{P}{M} \text{$\sigma_{\tmop{eff}}
  (\tmmathbf{P}, i)$}, \nonumber
\end{eqnarray}
where the l.h.s. denotes the classical loss term appearing in
Eq.~(\ref{eq:classicaleq}), one has for an ideal gas
\begin{eqnarray}
  \Gamma_{\tmmathbf{P}_0}^{i i} & = & n_{\tmop{gas}} \frac{P_0}{M}
  \sigma_{\tmop{eff}} (\tmmathbf{P}_0, i) = \frac{p}{Mk_{\text{B}} T} P_0
  \sigma_{\tmop{eff}} (\tmmathbf{P}_0, i),  \label{eq:eff}
\end{eqnarray}
where $p$ is the pressure of the gas and $T$ its temperature. One can thus
introduce a family of reference pressures depending on the initial internal
state of the particle entering in the interferometer
\begin{eqnarray}
  p_0^i & = & \frac{M k_{\text{B}} T}{P_0 \sigma_{\tmop{eff}} (\tmmathbf{P}_0,
  i) t},  \label{eq:poi}
\end{eqnarray}
where $t$ is the time of flight, so that Eq.~(\ref{eq:vvt}) can equivalently
be written as a function of the pressure in the interferometer
\begin{eqnarray}
  \Gamma_{\tmmathbf{P}_0}^{i i} t & = & \frac{p}{p^i_0} .  \label{eq:press}
\end{eqnarray}
This simply implies that the behaviour of the visibility as a function of time
is equivalent to its behaviour with respect to the pressure and therefore the
disturbance of the environment.

\section{Conclusions\label{sec:ceo}}

We have derived the master equation describing the dynamics of a test particle
with both translational and internal degrees of freedom, interacting through
collisions with a low density background gas. This has been done building on
the so-called monitoring approach {\cite{Hornberger2007b}}, and confirms a
previous heuristic argument put forward by one of us in
{\cite{Vacchini2008a}}. The present microscopic derivation further allows to
determine the energy shift. As we have checked, the result reduces to known
equations in suitable limits: the quantum linear Boltzmann equation if the
internal degrees of freedom are neglected {\cite{Vacchini2009a}}, the master
equation for an immobile test particle if the translational degrees of freedom
are not relevant {\cite{Dumcke1985a,Hornberger2007b}}, as well as quantum
classical Markovian master equations if one or both kind of degrees of freedom
can be treated as classical. Note that the natural bases in the derivation was
given by momentum for the motional degrees of freedom and energy for the
internal ones, the latter corresponding to the channel basis of scattering
theory. In these cases different channels are only coupled through the
collision term. If another internal basis can be of interest, also coherent
tunnelling effects appear.\\
We have further focused on the situation in which the internal degrees of
freedom, in spite of influencing the collisional scattering cross section, are
not probed by the measuring apparatus and therefore have to be averaged out
from the set of the observed dynamical variables, thus effectively becoming
part of the environment. The equation obtained in this situation is no more of
Lindblad type, but rather takes the form of a generalized Lindblad structure
{\cite{Breuer2007a,Vacchini2008a}}. It can therefore describe behaviour quite
different from that characterizing a Markovian dynamics. Solving these
equations in the position representation for an initial factorized state, we
have obtained an explicit expression for the visibility reduction in
interferometric experiments when internal degrees of freedom are involved. The
behaviour of the visibility can indeed be quite different from the exponential
decay corresponding to a Markovian dynamics, showing up e.g. oscillations and
revivals.\\
The interplay between different degrees of freedom in a bipartite system is a
natural source of non-Markovian behaviour, when either degrees of freedom
cannot be controlled, thus acting as an environment. This scenario has here
been studied in a concrete setting, assuming a factorized initial state, and
describing the dynamics in terms of a generalized Lindblad structure. Such a
choice of initial condition is relevant for the considered interferometric
setting, it would be however of great importance to consider initially
correlated states, which naturally appear when considering a non-Markovian
dynamics in a strong coupling regime. It would further be of interest to study
whether, instead of using this generalized Lindblad structure, one can obtain
a closed description of the reduced system dynamics in terms of a master
equation with a memory kernel, at least in some simplified situations. We plan
to address these topics in future research work.

\section{Acknowledgments}

We are grateful to Ludovico Lanz for useful hints along the preparation of the
manuscript. BV would also like to thank Klaus Hornberger for very helpful
discussions and reading of the manuscript. This work was partially supported
by MIUR under PRIN2008.

\appendix

\section{Microscopic derivation of the master equation}
\label{sec:micr-deriv-mast}

We here address the derivation of the Markovian master equation for
the description of the dynamics of the test particle starting from the
general expressions Eq.~\eqref{eq:ms} and Eq.~\eqref{eq:r}.
The scattering and rate operators appearing in these equations are best
expressed using the factorization of the total Hilbert space
$\mathcal{H}_{\tmop{tot}} = \mathcal{H_{\tmop{sys}} \otimes
\mathcal{H_{\text{gas}} =}} \mathcal{H_{\text{cm}} \mathcal{\otimes
H_{\text{rel}}}}$, according to
\begin{eqnarray}
 |\tmmathbf{P}, i \rangle \langle \tmmathbf{P}', j| \otimes |\tmmathbf{p}
  \rangle \langle \tmmathbf{p}' |_{\tmop{gas}} & = &| \tmop{rel} (\tmmathbf{p}, \tmmathbf{P}), i \rangle
  \langle \tmop{rel} (\tmmathbf{p}', \tmmathbf{P}'), j|_{\tmop{rel}}
 \otimes |\tmmathbf{P}+\tmmathbf{p} \rangle \langle \tmmathbf{P}' +\tmmathbf{p}'|_{\tmop{cm}} ,
  \label{eq:rel}
\end{eqnarray}
where the Hilbert space $\mathbbm{C}^n$ associated to the internal
degrees of freedom is part of $\mathcal{H}_{\text{rel}}$
{\cite{Taylor1972a}}, the notation is the same as in
Sec.~\ref{sec:ms}.  In fact both these operators act in a trivial way
on centre of mass coordinates: $\mathsf{\Gamma} =
\mathsf{I}_{\tmop{cm}} \otimes \mathsf{\Gamma}_0$ and $\mathsf{T} =
\mathsf{I}_{\tmop{cm}} \otimes \mathsf{T}_0$. The operator
$\mathsf{\Gamma}_0$ is given by
\begin{eqnarray}
  \mathsf{\Gamma_0} & = & \frac{n_{\text{gas}}}{m_{\ast}} \sum_j | \text{rel}
  ( \mathsf{p}, \mathsf{P}) | \sigma_{\text{tot}} ( \text{rel} ( \mathsf{p},
  \mathsf{P}), j) \otimes |j \rangle \langle j|,  \label{eq:gamma}
\end{eqnarray}
where $\sigma_{\text{tot}} ( \text{rel} (\tmmathbf{p}, \tmmathbf{P}), j)$ is the
total cross section, depending on initial relative momentum and internal
state.
The relation {\cite{Taylor1972a}}
\begin{equation}\label{eq:t}
    \langle \tmmathbf{p}_f, k| \mathsf{T_0} |\tmmathbf{p}_i, j \rangle =
  \frac{1}{2 \pi \hbar m_{\ast}} \delta \left( \frac{\tmmathbf{p}^2_f
  -\tmmathbf{p}^2_i}{2 m_{\ast}} +\mathcal{E}_{k j} \right) f_{k j}
  (\tmmathbf{p}_f, \tmmathbf{p}_i)
\end{equation}
links the operator $\mathsf{T}_0$ to the multichannel complex
scattering amplitudes $f_{k j} (\tmmathbf{p}_f, \tmmathbf{p}_i)$,
referring to scattering from an initial momentum $\tmmathbf{p}_i$ and
internal state $j$ to a final state with momentum $\tmmathbf{p}_f$ and
internal state $k$.  According to standard usage in scattering theory
we call channels the asymptotically free internal energy
eigenstates of the system. The differential cross section is given by
$\sigma_{k j} (\tmmathbf{p}_f, \tmmathbf{p}_i) = ({|\tmmathbf{p}_f
  |}/{|\tmmathbf{p}_i |}) | f_{k j} (\tmmathbf{p}_f, \tmmathbf{p}_i)
|^2$, so that the total cross section appearing in
Eq.~(\ref{eq:gamma}) reads $\sigma_{\tmop{tot}} (\tmmathbf{p}_i, j) =
\sum_k \int d\tmmathbf{p}_f \sigma_{k j} (\tmmathbf{p}_f,
\tmmathbf{p}_i)$.

\subsection{Evaluation of the Lindblad structure in momentum and internal
state basis}

\subsubsection{Incoherent contribution}
\label{incoh}

We now first concentrate on the evaluation of the contribution given
by Eq.~(\ref{eq:ms}), which under suitable approximations can be cast
in Lindblad form, closely following {\cite{Hornberger2008a}}, where
the special case of a test particle without internal structure was
dealt with. To this end we consider the matrix elements of
$\mathcal{L} \rho$ in the momentum and channel basis $\left.
   \{|\tmmathbf{P}, i \rangle \right\}$ of the Hilbert space $L^2
(\mathbbm{R}^3) \otimes \mathbbm{C}^n$ associated to the test
particle.  Denoting by $\mu (\tmmathbf{p})$ the stationary gas
momentum distribution and exploiting the relations Eq.~(\ref{eq:gamma}) and
Eq.~(\ref{eq:t}) for the relevant operators we can express the result
as is Eq.~\eqref{eq:rif2}, where the complex rate functions
\begin{eqnarray}
 \nonumber M^{j l}_{i k} \left( \tmmathbf{P}, \tmmathbf{P}', \tmmathbf{Q} \right) &
  \assign & \frac{(2 \pi \hbar)^3}{| \Omega |} \int d\tmmathbf{p}_0 \mu
  (\tmmathbf{p}_0)\langle \text{rel} \left( \tmmathbf{p}_0 -\tmmathbf{Q},
  \tmmathbf{P} \right), i| \mathsf{T}_0  \mathsf{\Gamma}^{1 / 2}_0 |
  \text{rel} (\tmmathbf{p}_0, \tmmathbf{P}-\tmmathbf{Q}), j \rangle
  \nonumber\\
  &  &\times\langle \text{rel} \left( \tmmathbf{p}_0, \tmmathbf{P}'
  -\tmmathbf{Q} \right)| \mathsf{\Gamma}^{1 / 2}_0 \mathsf{T}^{\dag}_0
  \text{rel} (\tmmathbf{p}_0 -\tmmathbf{Q}, \tmmathbf{P}')\rangle
  \label{eq:rate}
\end{eqnarray}
have been introduced, and $| \Omega |$ denotes the volume in which the gas is
confined. Note that $M^{jj}_{i i} (\tmmathbf{P}, \tmmathbf{P}; \tmmathbf{Q})$
can be interpreted as classical rates for scattering of the test particle with
momentum $\tmmathbf{P}-\tmmathbf{Q}$ and internal energy eigenstate $j$ to a
final state with momentum $\tmmathbf{P}$ and internal energy eigenstate $i$.
The contributions at r.h.s. of Eq.~(\ref{eq:rif2}) therefore play the role of
quantum gain and loss term also depending on the internal degrees of freedom
involved.

Relying on Eq.~(\ref{eq:rif2}) we can now deal just with the complex rates
$M^{j l}_{i k} (\tmmathbf{P}, \tmmathbf{P}' ; \tmmathbf{Q})$ defined in
Eq.~(\ref{eq:rate}). To proceed, it is helpful to introduce the following
functions of $\tmmathbf{p}_0$
\begin{eqnarray}
 \nonumber \tmmathbf{p}_i   & = &\text{rel} \left( \tmmathbf{p}_0,
  \frac{\tmmathbf{P}+\tmmathbf{P}'}{2} -\tmmathbf{Q} \right)\\
  \tmmathbf{p}_f &=& \text{rel} \left( \tmmathbf{p}_0 -\tmmathbf{Q},
  \frac{\tmmathbf{P}+\tmmathbf{P}'}{2} \right),  \label{eq:ppi-ppf}
\end{eqnarray}
which denote the mean of the pairs of initial and final relative momenta
appearing in $M^{j l}_{i k} (\tmmathbf{P}, \tmmathbf{P}' ; \tmmathbf{Q})$ and
which are related by $\tmmathbf{p}_i -\tmmathbf{p}_f =\tmmathbf{Q}$.
Introducing also $\tmmathbf{q}= \text{rel} \left( 0, \left(
\tmmathbf{P}-\tmmathbf{P}' \right) / 2 \right)$ the complex functions $M^{j
l}_{i k} \left( \tmmathbf{P}, \tmmathbf{P}' ; \tmmathbf{Q} \right)$ can be
expressed as an average over the gas distribution function $\mu$ of a complex
density in the centre of mass frame
\begin{eqnarray}
  M^{j l}_{i k} \left( \tmmathbf{P}, \tmmathbf{P}' ; \tmmathbf{Q} \right) & =
  & \int d\tmmathbf{p}_0 \mu (\tmmathbf{p}_0) m^{j l}_{i k} \left(
  \tmmathbf{p}_f, \tmmathbf{p}_i ; \tmmathbf{q} \right)  \label{eq:Mijkl}
\end{eqnarray}
with
\begin{eqnarray}
  \nonumber m^{j l}_{i k} \left( \tmmathbf{p}_f, \tmmathbf{p}_i ; \tmmathbf{q} \right) &
  = & \frac{(2 \pi \hbar)^3}{| \Omega |} \langle \tmmathbf{p}_f +\tmmathbf{q},
  i| \mathsf{T}_0 \mathsf{\Gamma}^{1 / 2}_0 |\tmmathbf{p}_i +\tmmathbf{q}, j
  \rangle \\
  &&\times \langle \tmmathbf{p}_i -\tmmathbf{q}, l| \mathsf{\Gamma}^{1 / 2}_0
  \mathsf{T}^{\dag}_0 |\tmmathbf{p}_f -\tmmathbf{q}, k \rangle .
  \label{eq:mijkl}
\end{eqnarray}
Evaluating this formula with the expressions Eq.~(\ref{eq:gamma}) and
Eq.~(\ref{eq:t}) for $\mathsf{\Gamma}_0$ and $\mathsf{T}_0$ respectively, we
obtain
\begin{eqnarray}
  \nonumber m^{j l}_{i k} \left( \tmmathbf{p}_f, \tmmathbf{p}_i ; \tmmathbf{q} \right) &
  = & \frac{2 \pi \hbar}{m_{\ast} | \Omega |} \mathsf{\Gamma}^{1 / 2}_0
  (\tmmathbf{p}_i +\tmmathbf{q}, j) \mathsf{\Gamma}^{1 / 2}_0 (\tmmathbf{p}_i
  -\tmmathbf{q}, j')\\
  & &\times  \delta \left(
  \frac{\tmmathbf{p}^2_f -\tmmathbf{p}^2_i}{2 m_{\ast}} -
  \frac{(\tmmathbf{p}_f -\tmmathbf{p}_i) \cdot \tmmathbf{q}}{m_{\ast}}
  +\mathcal{E}_{k l} \right)\nonumber\\
  &  & \nonumber \times \delta \left( \frac{\tmmathbf{p}^2_f -\tmmathbf{p}^2_i}{2
  m_{\ast}} + \frac{(\tmmathbf{p}_f -\tmmathbf{p}_i) \cdot
  \tmmathbf{q}}{m_{\ast}} +\mathcal{E}_{i j} \right)\\
 & &\times f_{i j} (\tmmathbf{p}_f +\tmmathbf{q}, \tmmathbf{p}_i
  +\tmmathbf{q}) f^{\ast}_{k l} (\tmmathbf{p}_f -\tmmathbf{q}, \tmmathbf{p}_i
  -\tmmathbf{q}) ,  \label{eq:m}
\end{eqnarray}
where $\Gamma_0 (\tmmathbf{p}_i, j) = n_{\tmop{gas}} |\tmmathbf{p}|
\sigma_{\tmop{tot}} (\tmmathbf{p}, j) / m_{\ast}$ is the eigenvalue of the
operator $\mathsf{\Gamma}_0$ relative to the state $|\tmmathbf{p}, j \rangle$.

The expression given by Eq.~(\ref{eq:m}) looses its meaning in the infinite
volume limit, due to appearance of the arbitrarily large normalization volume.
This point has been extensively discussed in {\cite{Hornberger2008a}}. It is
to be traced back to the fact that the operator $\mathsf{\Gamma}$ in order to
provide the actual rate of collisions should involve a projection on the
subspace of incoming wave packets, which is not accounted for in
Eq.~(\ref{eq:mijkl}). To do this, we are going now to evaluate the operator
$\mathsf{\Gamma}$ on a properly modified state of the relative motion.

Before that, it is convenient to focus our attention on the two
delta-functions appearing in Eq.~(\ref{eq:m}): employing the relation $\delta
(a) \delta (b) = 2 \delta (a + b) \delta (a - b)$, we can rewrite them as the
product
\begin{eqnarray*}
   \frac{1}{2} \delta \left( \frac{\tmmathbf{p}^2_f -\tmmathbf{p}^2_i}{2
  m_{\ast}} + \frac{\mathcal{E}_{i j} +\mathcal{E}_{k l}}{2} \right) \delta
  \left(  \frac{(\tmmathbf{p}_f -\tmmathbf{p}_i) \cdot \tmmathbf{q}}{m_{\ast}}
  + \frac{\mathcal{E}_{i j} -\mathcal{E}_{k l}}{2} \right) .
\end{eqnarray*}
These two constraints ensure that the scattering amplitudes appearing in
Eq.~(\ref{eq:m}) are evaluated on shell. The function $m^{j l}_{i k} \left(
\tmmathbf{p}_f, \tmmathbf{p}_i ; \tmmathbf{q} \right)$ gives a significant
contribution to the integral in Eq.~(\ref{eq:Mijkl}) when the two energy
differences are approximately equal, so that $\mathcal{E}_{i j}
=\mathcal{E}_{k l}$, leading otherwise to rapidly oscillating phases, and this
is actually a necessary condition in order to obtain a completely positive
time evolution {\cite{Alicki2003a}}. This implies in particular that
integrating the generalized function $m^{j l}_{i k} \left( \tmmathbf{p}_f,
\tmmathbf{p}_i ; \tmmathbf{q} \right)$ with a function $g (\tmmathbf{q})$, the
contributions deriving from the parallel component of $\tmmathbf{q}$ vanish
\begin{eqnarray}
  m^{j l}_{i k} \left( \tmmathbf{p}_f, \tmmathbf{p}_i ; \tmmathbf{q} \right) g
  (\tmmathbf{q}) & = & m^{j l}_{i k} \left( \tmmathbf{p}_f, \tmmathbf{p}_i ;
  \tmmathbf{q}_{\perp} \right) g (\tmmathbf{q}_{\perp}) .  \label{eq:qperp}
\end{eqnarray}
We now therefore evaluate $m^{j l}_{i k} \left( \tmmathbf{p}_f, \tmmathbf{p}_i
; \tmmathbf{q}_{\perp} \right)$ with a properly modified state of relative
motion, which takes into account the restriction of the expression to states
which actually describe a colliding pair. To this end we write the complex
rate $m^{j l}_{i k}$ as
\begin{eqnarray}
 \nonumber m^{j l}_{i k} \left( \tmmathbf{p}_f, \tmmathbf{p}_i ; \tmmathbf{q}_{\perp} \right) &
  = & \text{$\langle \tmmathbf{p}_f +\tmmathbf{q}_{\perp}, i| \mathsf{T}_0
  \mathsf{\Gamma}^{1 / 2}_0 \exp \left( i \frac{\mathsf{x}_{\tmop{rel}} \cdot
  \tmmathbf{q}_{\perp}}{\hbar} \right) \rho_{\tmmathbf{p}_i}$}\otimes |j
  \rangle \langle l|\\
  & & \times \text{$\exp \left( i \frac{\mathsf{x}_{\tmop{rel}} \cdot
  \tmmathbf{q}_{\perp}}{\hbar} \right) \mathsf{\Gamma}^{1 / 2}_0
  \mathsf{T}^{\dag}_0 |\tmmathbf{p}_f -\tmmathbf{q}_{\perp}, k \rangle$} \nonumber
\end{eqnarray}
where $\rho_{\tmmathbf{p}_i}$ denotes an improper state of relative motion
\begin{eqnarray}
  \rho_{\tmmathbf{p}_i} & = & \frac{(2 \pi \hbar)^3}{| \Omega |}
  |\tmmathbf{p}_i \rangle \langle \tmmathbf{p}_i |.  \label{eq:improper}
\end{eqnarray}
Since the rate operator $\mathsf{\Gamma}$ should have a vanishing expectation
value for those states of the relative motion that are not of the incoming
type, we make the replacement
\begin{eqnarray}
  \nonumber \rho_{\tmmathbf{p}_i } \otimes |j \rangle \langle l|& \rightarrow&
  \rho_{\tmmathbf{p}_i}' \otimes |j \rangle \langle l|  =
  \int_{\Lambda_{\tmmathbf{p}_i}}  \frac{d\tmmathbf{x}_{\parallel
  \tmmathbf{p}_i}}{| \Lambda_{\tmmathbf{p}_i} |}
  \int_{\Sigma_{\tmmathbf{p}_i}}  \frac{d\tmmathbf{x}_{\perp
  \tmmathbf{p}_i}}{| \Sigma_{\tmmathbf{p}_i} |} \int
  d\tmmathbf{w}e^{i\tmmathbf{x} \cdot \tmmathbf{w}/ \hbar} |\tmmathbf{p}_i -
  \frac{\tmmathbf{w}}{2}, j \rangle \langle \tmmathbf{p}_i +
  \frac{\tmmathbf{w}}{2}, l|. \\ \label{newstate}
\end{eqnarray}
This corresponds to a restriction of the Wigner function associated to
the improper state of relative motion Eq.~(\ref{eq:improper}) from the
entire normalization volume $| \Omega |$ to a cylinder pointing in the
direction $\tmmathbf{p}_i$, with base surface
$\Sigma_{\tmmathbf{p}_i}$ and height $\Lambda_{\tmmathbf{p}_i}$. As in
the case without internal degrees of freedom
$\Lambda_{\tmmathbf{p}_i}$ is approximately the distance travelled by
the particle between two subsequent collisions, while
$\Sigma_{\tmmathbf{p}_i}$ can now be taken as the geometric mean of
the total cross-section of the involved channels
{\cite{Hornberger2007b}}, that is to say $ | \Sigma_{\tmmathbf{p}_i} |
= \sqrt{\sigma (\tmmathbf{p}_i, j) \sigma (\tmmathbf{p}_i, l)}$ and $|
\Lambda_{\tmmathbf{p}_i} | = ({|\tmmathbf{p}_i |}\Delta t) / {m_{\ast}
}$, with $\Delta t$ the typical time interval between two subsequent collisions.

Putting the new state Eq.~(\ref{newstate}) into the equation
Eq.~(\ref{eq:mijkl}) and using the expressions of the matrix elements of
$\mathsf{T}_0$ and $\mathsf{\Gamma}_0$, we get
\begin{widetext}
\begin{eqnarray}
  m^{j l}_{i k} \left( \tmmathbf{p}_f, \tmmathbf{p}_i ; \tmmathbf{q}_{\perp}
  \right) & = & \int_{\Lambda_{\tmmathbf{p}_i}}
  \frac{d\tmmathbf{x}_{\parallel \tmmathbf{p}_i}}{| \Lambda_{\tmmathbf{p}_i}
  |} \int_{\Sigma_{\tmmathbf{p}_i}}  \frac{d\tmmathbf{x}_{\perp
  \tmmathbf{p}_i}}{| \Sigma_{\tmmathbf{p}_i} |} \int d\tmmathbf{w} \exp \left(
  - i \frac{\tmmathbf{x} \cdot \tmmathbf{w}}{\hbar} \right)  \frac{1}{(2 \pi
  \hbar m_{\ast})^2} \frac{n_{\tmop{gas}}}{|\tmmathbf{p}_i |} \nonumber\\
  &  & \times \delta \left( \frac{\tmmathbf{p}^2_f -\tmmathbf{p}^2_i}{2
  m_{\ast}} - \frac{\tmmathbf{w}^2}{8 m_{\ast}} - \frac{\tmmathbf{q}_{\perp}
  \cdot \tmmathbf{w}}{2 m_{\ast}} + \frac{\mathcal{E}_{i j} +\mathcal{E}_{k
  l}}{2} \right) \delta \left( \frac{\tmmathbf{p}_i \cdot
  \tmmathbf{w}}{|\tmmathbf{p}_i |} + \frac{m_{\ast}}{|\tmmathbf{p}_i |}
  (\mathcal{E}_{k l} -\mathcal{E}_{i j}) \right) \nonumber\\
  &  & \times f_{i j}  \left( \tmmathbf{p}_f +\tmmathbf{q}_{\perp},
  \tmmathbf{p}_i +\tmmathbf{q}_{\perp} + \frac{\tmmathbf{w}}{2} \right)
  f^{\ast}_{k l} \left( \tmmathbf{p}_f -\tmmathbf{q}_{\perp}, \tmmathbf{p}_i
  -\tmmathbf{q}_{\perp} - \frac{\tmmathbf{w}}{2} \right) \nonumber\\
  &  & \times \sqrt{\left| \tmmathbf{p}_i +\tmmathbf{q}_{\perp} +
  \frac{\tmmathbf{w}}{2} \right| \left| \tmmathbf{p}_i -\tmmathbf{q}_{\perp} -
  \frac{\tmmathbf{w}}{2} \right|} \nonumber\\
  & &\times  \sqrt{\sigma \left( \tmmathbf{p}_i
  +\tmmathbf{q}_{\perp} + \frac{\tmmathbf{w}}{2}, j \right) \sigma \left(
  \tmmathbf{p}_i -\tmmathbf{q}_{\perp} - \frac{\tmmathbf{w}}{2}, l \right)},
  \nonumber
\end{eqnarray}
\end{widetext}
where we have exploited once again the relation $\delta (a) \delta (b) = 2
\delta (a + b) \delta (a - b)$. We now first perform the integral over
$\tmmathbf{w}_{\parallel \tmmathbf{p}_i}$, which denotes the component of
$\tmmathbf{w}$ parallel to $\tmmathbf{p}_i$, thus evaluating the second
delta-function at r.h.s. of the previous equation, so that the dependence on
$\tmmathbf{x}_{\parallel \tmmathbf{p}_i}$ only appears in the term
\begin{eqnarray*}
  &  & \int_{\Lambda_{\tmmathbf{p}_i}}  \frac{d\tmmathbf{x}_{\parallel
  \tmmathbf{p}_i}}{| \Lambda_{\tmmathbf{p}_i} |} \exp \left( - \frac{i}{\hbar}
  \tmmathbf{x}_{\parallel \tmmathbf{p}_i} \cdot \left(
  \frac{m_{\ast}}{|\tmmathbf{p}_i |} (\mathcal{E}_{i j} -\mathcal{E}_{k l})
  \right) \hat{\tmmathbf{p}}_i \right) .
\end{eqnarray*}
The phase of the integrand varies very quickly for $(\mathcal{E}_{i j}
-\mathcal{E}_{k l}) \gg \hbar / \Delta t$, where $\Delta t$ is the typical
time elapsing between collisions, so that as already discussed its
contribution vanishes unless $\mathcal{E}_{i j} =\mathcal{E}_{k l}$,
corresponding to a rotating wave approximation, assuming a separation of time
scales between internal and translational dynamics {\cite{Breuer2007}}.
Further considering the integral over $\tmmathbf{x}_{\perp \tmmathbf{p}_i}$ as
an approximate expression for $\delta \left( \tmmathbf{w}_{\perp
\tmmathbf{p}_i} \right)$ we are led to
\begin{eqnarray}
\nonumber  m^{j l}_{i k} \left( \tmmathbf{p}_f, \tmmathbf{p}_i ; \tmmathbf{q}_{\perp}
  \right)  & = & \frac{n_{\tmop{gas}}}{m_{\ast}^2} \chi^{jl}_{ik}  f_{i j}  \left( \tmmathbf{p}_f +\tmmathbf{q}_{\perp},
  \tmmathbf{p}_i +\tmmathbf{q}_{\perp} \right)
\nonumber\\&&\times f^{\ast}_{k l} \left( \tmmathbf{p}_f -\tmmathbf{q}_{\perp},
  \tmmathbf{p}_i -\tmmathbf{q}_{\perp} \right) \delta
  \left( \frac{\tmmathbf{p}^2_f -\tmmathbf{p}^2_i}{2 m_{\ast}} +\mathcal{E}_{i
  j} \right) \nonumber\\
  \nonumber & &\times \frac{\sqrt{|\tmmathbf{p}_i +
  \tmmathbf{q}_{\perp} | |\tmmathbf{p}_i -\tmmathbf{q}_{_{\perp}}
  |}}{|\tmmathbf{p}_i |} \frac{ \sqrt{\sigma \left( \tmmathbf{p}_i
  +\tmmathbf{q}_{\perp}, j \right) \sigma \left( \tmmathbf{p}_i
  -\tmmathbf{q}_{\perp}, l \right)}}{ \sqrt{\sigma (\tmmathbf{p}_i, j) \sigma
  (\tmmathbf{p}_i, l)}}, \nonumber
\end{eqnarray}
where the $\chi^{jl}_{ik}$ act like a Kronecker's delta factor, being
defined according to
\begin{eqnarray}
  \chi^{jl}_{ik} & = & \left\{ \begin{array}{ll}
    1 & \tmop{if} \mathcal{E}_{i j} =\mathcal{E}_{k l}\\
    0 & \tmop{otherwise}
  \end{array} \right. . \hspace*{\fill} \nonumber
\end{eqnarray}

In the last two terms we can disregard the dependence on
$\tmmathbf{q}_{\perp}$ because we expect that a
$\tmmathbf{q}_{\perp}$-integration will average out the ``far off-diagonal''
contributions with large modulus $|\tmmathbf{q}_{\perp} |$, where the phases
of the two scattering amplitudes are no longer synchronous. In conclusion, we
have
\begin{eqnarray}
  m^{j l}_{i k} \left( \tmmathbf{p}_f, \tmmathbf{p}_i ; \tmmathbf{q}_{\perp}
  \right) & = & \frac{n_{\tmop{gas}}}{m_{\ast}^2} \chi^{jl}_{ik} \delta \left(
  \frac{\tmmathbf{p}^2_f -\tmmathbf{p}^2_i}{2 m_{\ast}} + E_i - E_j \right)
 \nonumber\\
 & & \times f_{i j}  \left( \tmmathbf{p}_f +\tmmathbf{q}_{\perp},
  \tmmathbf{p}_i +\tmmathbf{q}_{\perp} \right) f^{\ast}_{k l} \left(
  \tmmathbf{p}_f -\tmmathbf{q}_{\perp}, \tmmathbf{p}_i -\tmmathbf{q}_{\perp}
  \right) .  \label{eq:mris}
\end{eqnarray}
This relation determines the 
complex rate functions $M^{j l}_{i k} \left( \tmmathbf{P}, \tmmathbf{P}' ;
\tmmathbf{Q} \right)$ through Eq.~(\ref{eq:Mijkl}), and therefore the
dissipative part of the master equation according to
Eq.~(\ref{eq:rif2}).

\subsubsection{Energy shift}
\label{coh}

As a last step in the determination of the structure of the master equation we
need to evaluate the contribution given by $\mathcal{R} \rho$. In the same
notation as above, and within the same approximations, we directly obtain
\begin{eqnarray}
 \nonumber \langle \tmmathbf{P}, i| \mathcal{R} \rho |\tmmathbf{P}', k \rangle & = &
  \langle \tmmathbf{P}, i| i \text{ Tr}_{\tmop{gas}} \left( \left[
  \mathsf{\Gamma}^{1 / 2} \text{Re} ( \mathsf{T}) \mathsf{\Gamma}^{1 / 2}, \rho \otimes \rho_{\tmop{gas}} \right]  \right) |\tmmathbf{P}', k \rangle\\
 & = & \frac{1}{i \hbar} \sum_j \left( E^{i j}_n (\tmmathbf{P}) \langle
  \tmmathbf{P}, j| \rho |\tmmathbf{P}', k \rangle- E^{j k}_n (\tmmathbf{P}')
  \langle \tmmathbf{P}, i| \rho |\tmmathbf{P}', j \rangle \right),
  \label{eq:coherent}
\end{eqnarray}
with
\begin{eqnarray}
  E^{i j}_n (\tmmathbf{P}) & = & \text{$- 2 \pi \hbar^2
  \frac{n_{\tmop{gas}}}{m_{\ast}} \chi^{jk}_{ik} \int d\tmmathbf{p}_0 \mu
  (\tmmathbf{p}_0)$} \text{Re}  \left[ f_{i j} \left( \text{rel} \left(
  \tmmathbf{p}_0, \tmmathbf{P} \right), \text{rel} \left( \tmmathbf{p}_0,
  \tmmathbf{P} \right) \right) \right] .  \label{eq:Enij}
\end{eqnarray}
It is worth noting that for the case of a non-degenerate free internal
Hamiltonian this formula reduces to
\begin{eqnarray}
  \langle \tmmathbf{P}, i| \mathcal{R} \rho |\tmmathbf{P}', k \rangle & = &
  \frac{1}{i \hbar} \left( E^i_n (\tmmathbf{P}) - E^k_n (\tmmathbf{P}')
  \right) \langle \tmmathbf{P}, i| \rho |\tmmathbf{P}', k \rangle \nonumber
\end{eqnarray}
with
\begin{eqnarray}
  E^i_n (\tmmathbf{P}) & = & - 2 \pi \hbar^2 \frac{n_{\tmop{gas}}}{m_{\ast}}
  \int d\tmmathbf{p}_0 \mu (\tmmathbf{p}_0)\text{Re}  \left[ f_{i i} \left(
  \text{rel} \left( \tmmathbf{p}_0, \tmmathbf{P} \right), \text{rel} \left(
  \tmmathbf{p}_0, \tmmathbf{P} \right) \right) \right] .  \label{eq:eni}
\end{eqnarray}

\subsection{Operator expression of the master equation\label{sec:operator}}

We now recast the master equation Eq.~(\ref{eq:qlbefinal}),
whose matrix elements are given by Eq.~(\ref{eq:rif2}) and
Eq.~(\ref{eq:coherent}), in a way which allows to express it in a representation-independent form. 
The key point is
to show that $M^{j l}_{i k} \left( \tmmathbf{P}, \tmmathbf{P}' ; \tmmathbf{Q}
\right) = \int d\tmmathbf{p}_0 \mu (\tmmathbf{p}_0) m^{j l}_{i k} \left(
\tmmathbf{p}_f, \tmmathbf{p}_i ; \tmmathbf{q} \right)$ can be factorized into
two terms, one depending on $\tmmathbf{P}$ and the other on $\tmmathbf{P}'$.

Changing the integration variable from $\tmmathbf{p}_0$ to $\tmmathbf{p}_i$
and using the relations Eq.~(\ref{eq:ppi-ppf}) to obtain $\tmmathbf{p}_0
=\tmmathbf{p}_i + \left( \tmmathbf{p}_f +\tmmathbf{P} \right) m / M
+\tmmathbf{q}m / m_{\ast} =\tmmathbf{p}_i + (\tmmathbf{p}_f +\tmmathbf{P}') m
/ M -\tmmathbf{q}m / m_{\ast}$, we have
\begin{eqnarray}
\nonumber  M^{j l}_{i k} \left( \tmmathbf{P}, \tmmathbf{P}' ; \tmmathbf{Q} \right) & =
  & \frac{m^3}{m^3_{\ast}} \frac{n_{\tmop{gas}}}{m_{\ast}} \chi^{jl}_{ik} \int
  d\tmmathbf{p}_i  \delta \left( \frac{\tmmathbf{p}^2_f -\tmmathbf{p}^2_i}{2
  m_{\ast}} +\mathcal{E}_{i j} \right)  \mu^{1 / 2} \left( \tmmathbf{p}_i + \frac{m}{M} \left(
  \tmmathbf{p}_f +\tmmathbf{P} \right) + \frac{m}{m_{\ast}}
  \tmmathbf{q}_{\perp} \right) \nonumber\\
  &  & \times \mu^{1 / 2} \left( \tmmathbf{p}_i + \frac{m}{M} (\tmmathbf{p}_f
  +\tmmathbf{P}') - \frac{m}{m_{\ast}} \tmmathbf{q}_{\perp} \right)\nonumber\\
  & &\times  f_{i j}  \left( \tmmathbf{p}_f
  +\tmmathbf{q}_{\perp}, \tmmathbf{p}_i +\tmmathbf{q}_{\perp} \right)
  f^{\ast}_{k l} \left( \tmmathbf{p}_f -\tmmathbf{q}_{\perp}, \tmmathbf{p}_i
  -\tmmathbf{q}_{\perp} \right), \nonumber
\end{eqnarray}
where we replaced $\tmmathbf{q}$ by $\tmmathbf{q}_{\perp}$ in the arguments of
$\mu^{1 / 2}$, in accordance with Eq.~(\ref{eq:qperp}). Remembering that
$\tmmathbf{p}_i -\tmmathbf{p}_f =\tmmathbf{Q}$ and $\tmmathbf{q}= m_{\ast}
\left( \tmmathbf{P}-\tmmathbf{P}' \right) / \left( 2 M \right)$, we consider
the change of variable
\begin{eqnarray}
  \tmmathbf{p}_i & \rightarrow & \frac{m}{m_{\ast}} \tmmathbf{p}_i +
  \frac{m}{M} \frac{\tmmathbf{P}_{\perp} +\tmmathbf{P}'_{\perp}}{2} -
  \frac{m}{m_{\ast}} \frac{\tmmathbf{Q}}{2} - \frac{\mathcal{E}_{i j}}{Q^2 /
  m} \tmmathbf{Q}  \label{eq:newp}
\end{eqnarray}
to obtain the desired factorization. If we further consider that the
delta function $\delta \left( \tmmathbf{p} \cdot \tmmathbf{Q}/ m
\right)$ restricts the $\tmmathbf{p}$-integration to the plane
$\tmmathbf{Q}_{\perp} = \{ \tmmathbf{p} \in \mathbbm{R}^3 :
\tmmathbf{p} \cdot \tmmathbf{Q}= 0 \}$ we finally arrive at the
expression Eq.~\eqref{eq:fact}, with $L_{i j} \left( \tmmathbf{p},
   \tmmathbf{P}; \tmmathbf{Q} \right)$ as in Eq.~\eqref{eq:L}, which
allows to obtain the operator expression of the master equation given
by Eq.~(\ref{eq:diag}) and Eq.~(\ref{eq:hn}).  If the gas distribution
function $\mu$ is given by a Maxwell-Boltzmann probability density
$\mu_{\beta} (\tmmathbf{p}) = {1}/{(\pi^{3 / 2} p^3_{\beta})} \exp ( -
{\tmmathbf{p}^2}/{p^2_{\beta}})$, where $p_{\beta} = \sqrt{2 m /
  \beta}$ is the most probable momentum at temperature $T = 1 / (k_B
\beta)$, these functions can be expressed in terms of the dynamic
structure factor for a Maxwell-Boltzmann gas
{\cite{Schwabl2003,Pitaevskii2003}}
\begin{eqnarray}
  S_{\tmop{MB}} (\tmmathbf{Q}, E) & = & \sqrt{\frac{\beta m}{2 \pi}}
  \frac{1}{Q} \exp \left( - \frac{\beta}{8 m}  \frac{(Q^2 + 2 m E)^2}{Q^2}
  \right) .  \label{eq:smb}
\end{eqnarray}
In fact, using the relation
\begin{eqnarray*}
 & & \frac{m}{Q} \mu_{\beta} \left( \tmmathbf{p}_{\perp} + \frac{m}{M}
  \tmmathbf{P}_{\parallel} + \left( 1 + \frac{m}{M} \right)
  \frac{\tmmathbf{Q}}{2} + \frac{\mathcal{E}_{i j}}{Q^2 / m} \tmmathbf{Q}
  \right) = \mu_{\beta} (\tmmathbf{p}_{\perp}) S_{\tmop{MB}}
  (\tmmathbf{Q}, E (\tmmathbf{Q}, \tmmathbf{P}) + \mathcal{E}_{i j}),
\end{eqnarray*}
with $E (\tmmathbf{Q}, \tmmathbf{P}) = : {(\tmmathbf{P}+\tmmathbf{Q})^2}/{2
  M} - {\tmmathbf{P}^2}/{2 M}$
the energy transferred to the centre of mass in a collision changing the
momentum of the test particle from $\tmmathbf{P}$ to
$\tmmathbf{P}+\tmmathbf{Q}$, Eq.~(\ref{eq:L}) can be written as
{\cite{Vacchini2008a}}
\begin{eqnarray}
  L_{i j} \left( \tmmathbf{p}, \tmmathbf{P}; \tmmathbf{Q} \right) & = &
  \sqrt{\frac{n_{\tmop{gas}} }{m^2_{\ast}} \mu_{\beta} (\tmmathbf{p}_{\perp})}
  \sqrt{S_{\tmop{MB}} (\tmmathbf{Q}, E (\tmmathbf{Q}, \tmmathbf{P}) +
  \mathcal{E}_{i j})} \nonumber\\
  &  & \times f_{i j} \left( \text{rel} \left( \tmmathbf{p}_{\perp},
  \tmmathbf{P}_{\perp} \right) - \frac{\tmmathbf{Q}}{2} + \frac{\mathcal{E}_{i
  j}}{Q^2 / m_{\ast}} \tmmathbf{Q}, \text{rel} \left( \tmmathbf{p}_{\perp},
  \tmmathbf{P}_{\perp} \right) + \frac{\tmmathbf{Q}}{2} + \frac{\mathcal{E}_{i
  j}}{Q^2 / m_{\ast}} \tmmathbf{Q} \right) .\nonumber\\  \label{eq:L2}
\end{eqnarray}
In the latter expression for the Lindblad operators, the dynamic structure
factor appears evaluated for an energy transfer corresponding to the sum of
the contributions for centre of mass and internal state, as naturally
expected. As discussed in {\cite{Vacchini2009a}}, the dynamic structure factor
describes momentum and energy transferred to the test particle when scattering
off a macroscopic system, thus allowing for a more transparent physical
understanding of the structure of the Lindblad operators.

\end{document}